\documentclass[aps,prc,reprint,groupedaddress]{revtex4-1}
\usepackage{graphicx}
\usepackage{amsmath}
\begin{document}

% Use the \preprint command to place your local institutional report
% number in the upper righthand corner of the title page in preprint mode.
% Multiple \preprint commands are allowed.
% Use the 'preprintnumbers' class option to override journal defaults
% to display numbers if necessary
%\preprint{}

%Title of paper
\title{Near-barrier Photofission in $^{232}$Th and $^{238}$U}

% repeat the \author .. \affiliation  etc. as needed
% \email, \thanks, \homepage, \altaffiliation all apply to the current
% author. Explanatory text should go in the []'s, actual e-mail
% address or url should go in the {}'s for \email and \homepage.
% Please use the appropriate macro foreach each type of information

% \affiliation command applies to all authors since the last
% \affiliation command. The \affiliation command should follow the
% other information
% \affiliation can be followed by \email, \homepage, \thanks as well.
%\author{}
%\email[]{Your e-mail address}
%\homepage[]{Your web page}
%\thanks{}
%\altaffiliation{}
%\affiliation{}

\author{J.A. Silano}
\affiliation{Nuclear and Chemical Sciences Division, Lawrence Livermore National Laboratory, Livermore, CA 94550, USA}
\affiliation{Triangle Universities Nuclear Laboratory, Durham, North Carolina 27708, USA}
\affiliation{Department of Physics and Astronomy, University of North Carolina - Chapel Hill, Chapel Hill, North Carolina 27599, USA}
\email{silano1@llnl.gov}
\author{H.J. Karwowski}
\affiliation{Triangle Universities Nuclear Laboratory, Durham, North Carolina 27708, USA}
\affiliation{Department of Physics and Astronomy, University of North Carolina - Chapel Hill, Chapel Hill, North Carolina 27599, USA}

%Collaboration name if desired (requires use of superscriptaddress
%option in \documentclass). \noaffiliation is required (may also be
%used with the \author command).
%\collaboration can be followed by \email, \homepage, \thanks as well.
%\collaboration{}
%\noaffiliation

\date{\today}

\begin{abstract}
A study of photofission of $^{232}$Th and $^{238}$U was performed using quasi-monoenergetic, 
linearly-polarized $\gamma$-ray beams from the High Intensity $\gamma$-ray Source at Triangle Universities Nuclear Laboratory. 
The prompt photofission neutron polarization asymmetries, neutron multiplicities and the photofission 
cross sections were measured in the near-barrier energy range of 4.3-6.0 MeV. 
This data set constitutes the lowest energy measurements of those observables to date using 
quasi-monoenergetic photons. 
Large polarization asymmetries are observed in both nuclei, consistent with the E1 excitation as 
observed by another measurement of this kind made in a higher energy range. 
Previous experimental evidence of a deep third minimum in the $^{238}$U fission barrier has been identified as an accelerator-induced background.
\end{abstract}

% insert suggested PACS numbers in braces on next line
\pacs{}
% insert suggested keywords - APS authors don't need to do this
%\keywords{}

%\maketitle must follow title, authors, abstract, \pacs, and \keywords
\maketitle

\section{Introduction}

There has been much progress over the past 80 years in understanding the fission process but a 
complete model of the fission mechanism has yet to be established.  Recent advances in computing 
power have facilitated the possibility of a comprehensive microscopic description of the fission 
process. A complete and quantitatively  accurate microscopic fission model would significantly impact a 
number of fission applications which are currently of interest. The physical observables of fission such 
as the fission cross section, prompt neutron multiplicity, fragment masses and angular distributions are 
determined by the structure of the fission barrier - the potential energy surface that an excited nucleus 
must overcome to split apart. In order for any calculation to be able to predict those observables 
accurately, it must be able to predict the effective fission barrier. The barrier can not be directly 
measured and thus can only be inferred through measurements of these fission observables. 
              
Recent experimental
 \cite{PhysRevC.80.011301, KRASZNAHORKAY199915, CSATLOS2005175, PhysRevC.87.044321} 
 and theoretical \cite{PhysRevC.87.044308, PhysRevC.87.054327, PhysRevC.91.014321} 
 results have been in disagreement over the topic of a ``third minimum," or third well, in 
the potential energy surface (PES) describing the fission barrier of actinides. The barriers are typically 
presented by projecting the multidimensional PES along the quadrupole deformation of the nucleus, $
\beta$. The second and third wells are then seen as relative minima in the potential energy curve of 
this 1D model at $\beta$ values corresponding to spheroidal nuclear shapes with major to minor axis 
ratios of approximately 2:1 and 3:1, respectively. Owing to the deformations of the nuclear shapes, 
states in these potential minima are referred to as super- and 
hyper-deformed \cite{PhysRevC.87.044308}. 

Theoretical calculations of fission barriers in U and Th nuclei tend to predict shallow or non-existent 
third minima, especially in the heavier isotopes such as $^{232}$Th and $^{238}$U. Theoretical results 
by McDonnell \textit{et al.} \cite{PhysRevC.87.054327} highlight the role of shell corrections in the 
prominence of the third minima for Th and light U isotopes. These self-consistent calculations were 
performed using finite-temperature superfluid nuclear density functional theory with the Skyrme energy 
density functionals SkM$^{*}$ and UNEDF1, the latter being optimized for fission studies. A shallow 
third minimum was observed in the lightest Th and U isotopes, with at most a very shallow third 
minimum or shoulder in the $^{232}$Th and $^{232}$U barriers. The cause of the neutron number $N$ dependence 
is identified as a neutron shell effect that reduces the third outer barrier, making the third minimum 
increasingly shallow with increasing $N$. The calculations were limited to $^{226, 228, 
230, 232}$Th and $^{228, 230, 232, 234}$U, but the clear trend with increasing $N$ leads 
to a prediction of no third minimum in $^{238}$U. These results are at odds with the relativistic mean 
field (MDC-RMF) model calculations of Zhao \textit{et al.} \cite{PhysRevC.91.014321} which show a 
shallow, but well-formed 0.5 MeV and 1.11 MeV deep third minima in the fission barrier in $^{232}$Th and $^{238}$U, respectively.
 A shallow third minimum in $^{232}$Th is also supported by the 
macroscopic-microscopic calculations of Jachimowicz \textit{et al.} \cite{PhysRevC.87.044308}, using 
for the first time an eight-dimensional PES for that nucleus.  After a proper inclusion of the dipole 
deformation, the depth of the third minimum was determined to be about 0.36 MeV. The authors conclude that 
new experimental study dedicated to hyper-deformation in $^{232}$Th is essential for the 
understanding of the third minima in actinide nuclei. 

\begin{figure}[h]
\centering
\includegraphics[width=3.375in]{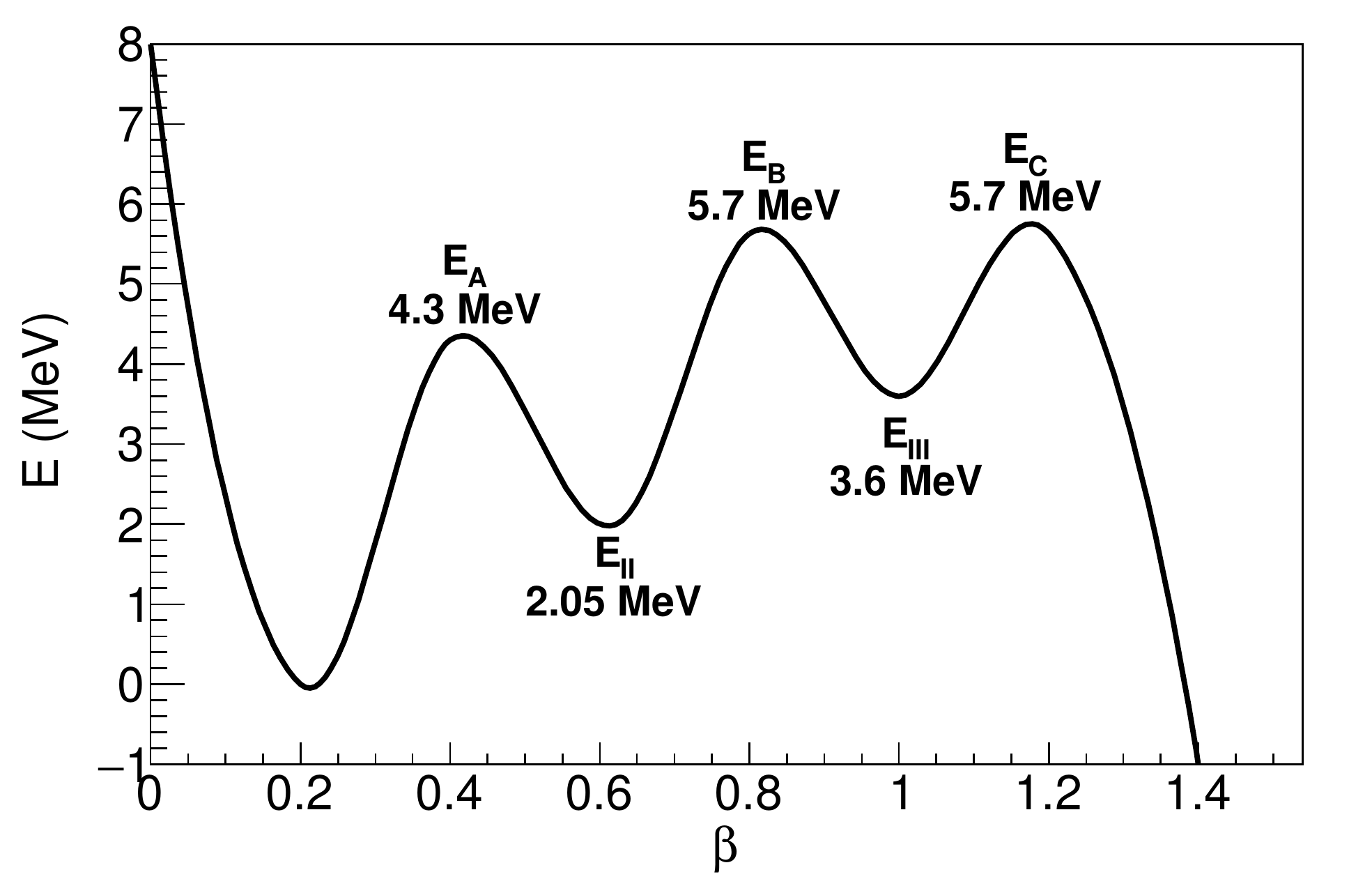}
\caption{A triple-humped interpretation of the $^{238}$U fission barrier based on barrier parameters from Ref. \cite{PhysRevC.87.044321}. The barrier parameters were determined by reproducing the experimentally measured photofission reaction cross section with statistical model calculations.}\label{fig:triple_humped_barrier}
\end{figure}

In contrast with shallow third minima favored by theoretical models, a new paradigm of the 
triple-humped fission barrier with a deep hyper-deformed third minimum for U and Th isotopes was 
developed from experimental data. When the double-humped fission barrier was initially established
\cite{vandenbosch1973nuclear,RevModPhys.52.725} for the lighter actinide isotopes, the so-called 
``Thorium anomaly" persisted. Strong transmission resonances were observed for Th nuclei, which 
pointed to equal heights of the inner and outer fission barrier \cite{PhysRevLett.28.1707}. However, the systematics of the fission 
barriers (and their theoretical predictions) suggested instead a much lower inner barrier than the outer 
one. In this context a triple-humped fission barrier was introduced assuming only a shallow third 
potential minimum. Almost two decades later it has been demonstrated with data from light-ion induced 
fission reactions that the outer third minimum for $^{232, 234, 236}$U is in fact as deep as the second 
minimum \cite{PhysRevC.80.011301, KRASZNAHORKAY199915, CSATLOS2005175}. Most recently, 
photofission cross section measurements performed by Csige \textit{et al.} 
 \cite{PhysRevC.87.044321} on $^{238}$U indicated a 2 MeV deep third well in the $^{238}$U
fission barrier as shown in Fig. \ref{fig:triple_humped_barrier}. However, while these data were interpreted to show pronounced preference for
triple-humped barrier in $^{238}$U, low statistics at near threshold energies did not allow for 
precise mapping of the resonance structure of the cross sections near threshold.

The experimental support for deep third minima in Th isotopes is more conflicted. Neutron-induced fission 
measurements by Blons \textit{et al.} \cite{BLONS19841, BLONS1988231, BLONS1989121} indicated 
a shallow third minimum of less than 0.5 MeV in the $^{230, 231, 233}$Th fission barriers. However a 
deep third minimum of $\sim2$ MeV in the $^{232}$Th fission barrier was obtained by Blokhin and 
Soldatov \cite{Blokhin2009} by analyzing photofission cross sections extracted from data obtained by unfolding 
bremsstrahlung beams \cite{Smirenkin:1996aa}. In light of the experimental evidence of deep third 
minima in the fission barriers of U isotopes, older $^{232}$Th photofission data
 \cite{PhysRevC.34.1397} was reinterpreted by Thirolf \textit{et al.} \cite{Thirolf_2012} to tentatively 
 show a $\sim4$ MeV deep third minimum, roughly equal in depth to the second minimum. The 
 prospects for further re-analysis of older $^{232}$Th photofission cross section data are limited by the 
 large discrepancies between data sets, particularly in the sub-barrier energy region where the effects 
 of the fission barrier structure are most pronounced \cite{Lindgren1978}.

Photofission, which is the tool used in the present work, has proven to be a valuable probe of the 
fission barrier structure since an incident  $\gamma$-ray photon brings a single unit of angular 
momentum into the fissioning system. Additionally, at energies near the fission barrier, the 
$\gamma$-ray photon interacts with the nucleus primarily through E1 transitions, greatly reducing the number of 
fission channels that contribute to the measured data. This is especially true for even-even nuclei such 
as $^{232}$Th and $^{238}$U, which have a $J^{\pi}=0^+$ ground state and thus can only be 
connected to a $1^-$ state by an E1 excitation. Recently, Mueller \textit{et al.} 
\cite{PhysRevC.89.034615} probed the spin and parity distribution of the fissioning compound nucleus 
in a number of actinides including $^{232}$Th and $^{238}$U, by measuring  polarization asymmetries 
in the angular distribution of prompt photofission neutrons induced by linearly polarized $\gamma$-ray 
beams. However these measurements were limited primarily to energies above the fission barrier.

Currently there are very few photofission data at low $\gamma$-ray energies where the effects of the 
fission barrier are most apparent, and there are significant discrepancies in some of the existing data 
sets. A majority of the available data comes from measurements performed using bremsstrahlung beams, meaning that the 
cross section data is dependent on the specific method of unfolding the beam spectrum applied to the measured 
fission yields. It is the aim of the present work to provide data that will help to better constrain 
the shape of the fission barrier through measurements of the photofission process on $^{232}$Th and 
$^{238}$U targets. Photofission was investigated with linearly polarized $\gamma$-ray beams with 
energies between 4.3 and 6.0 MeV, and prompt fission neutrons were detected to measure the 
photofission cross sections, photofission neutron polarization asymmetries and prompt fission neutron 
multiplicities. A majority of the existing photofission data was obtained using bremsstrahlung beams.  
The present  measurements were made  with quasi-monoenergetic  $\gamma$-ray beams thus avoiding 
 potential systematic error introduced by a beam energy deconvolution process.

\section{Experiment}
\subsection{$\gamma$-ray Beams}

Photofission measurements were performed at the High Intensity $\gamma$-ray Source 
(HI$\gamma$S), located at Triangle Universities Nuclear Laboratory (TUNL). The HI$\gamma$S 
facility \cite{Weller2009257} produces intense, quasi-monoenergetic, 100\% linearly polarized 
$\gamma$-ray beams by Compton back-scattering free electron laser (FEL) photons off of relativistic 
electrons. For this work $\gamma$-ray beams were produced in the range of 4.3-6.3 MeV with an 
energy resolution of $\leq 3\%$ and a typical flux on target of $\sim10^8$ $\gamma$/s. The accelerator 
was operated with 780 nm FEL photons, electron storage ring energies of $420-520$ MeV and a 
typical electron storage ring current of 90 mA. A 12 mm diameter, 15.24 cm long Pb collimator was used to limit the 
$\gamma$-ray beam size and define the energy resolution.

\begin{figure}[h]
\centering
\includegraphics[width=3.375in]{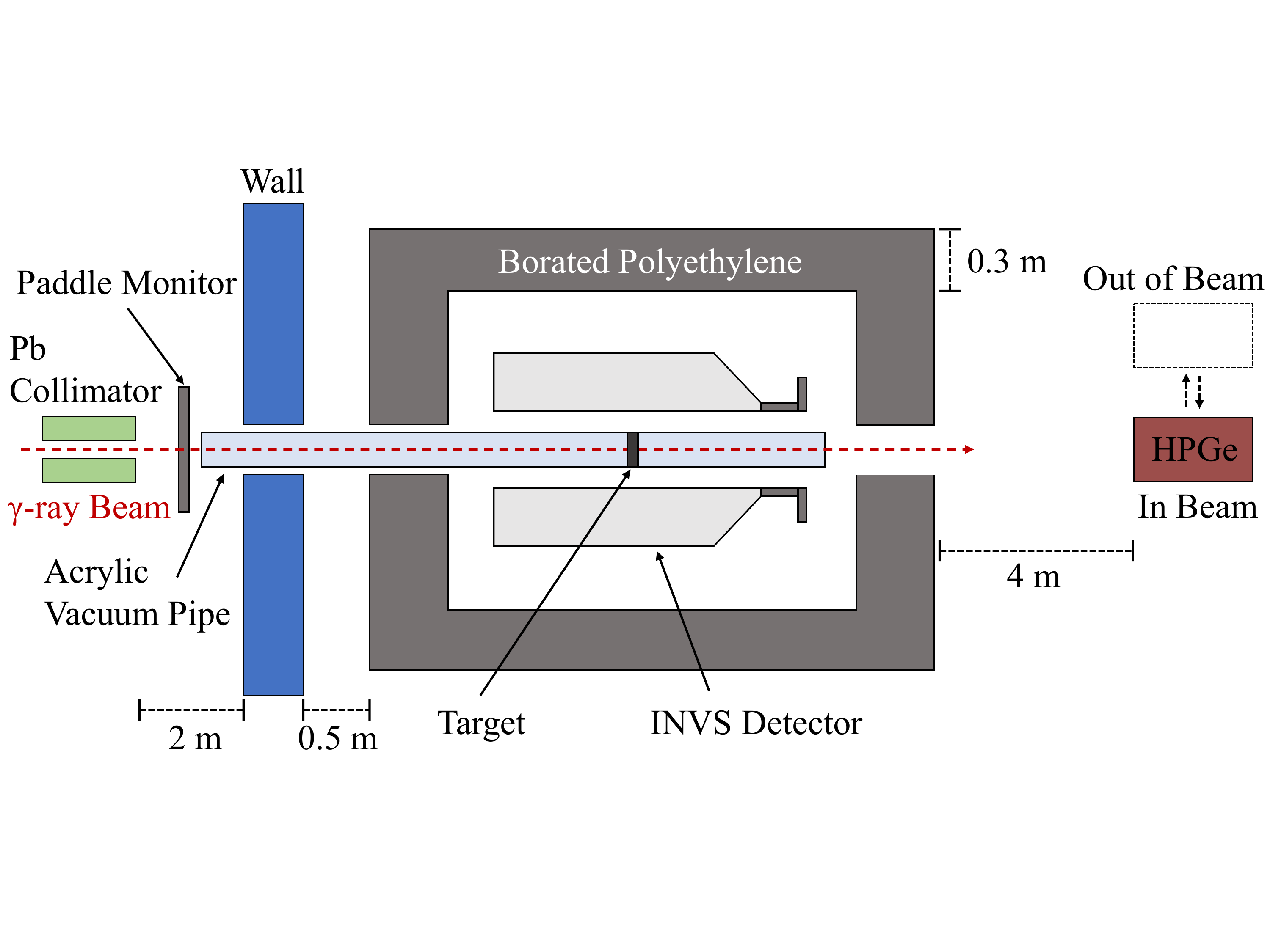}
\caption{(Color online) Illustration of the experimental geometry. The target was mounted in a vacuum pipe inserted through 
the center of the INVS neutron detector. Borated polyethylene neutron shielding surrounded the 
detector. An upstream scintillating paddle monitored the $\gamma$-ray beam flux, while a downstream 
HPGe could be moved into the beam to measure the energy spectrum.}\label{fig:experiment_overview}
\end{figure}

The experimental setup is shown in Fig. \ref{fig:experiment_overview}. The absolute flux of the 
$\gamma$-ray beam was measured placing a thin sheet of plastic scintillator upstream of the main 
experimental target and detector assembly. The flux monitor consisted of a 10 cm x 10 cm x 1 mm 
sheet of polyvinyltoluene affixed to a photomultiplier tube (PMT) by a light guide. Similar scintillating 
paddles were characterized as HI$\gamma$S $\gamma$-ray beam flux monitors by Pywell 
\textit{et al.} \cite{Pywell2009517}. The energy spectrum of the incident $\gamma$-ray beam was 
measured with a 120\% efficiency high purity germanium detector (HPGe). The HPGe was mounted 
on a movable platform allowing it to be positioned off-axis during the main photofission runs and in the 
$\gamma$-ray beam to measure the energy spectrum. In these measurements copper attenuators 
were placed upstream of the collimator to reduce the $\gamma$-ray beam flux and consequently the 
HPGe dead time to reasonable levels.

\subsection{Targets}

Table \ref{table:udisks} lists the properties of the $^{232}$Th, $^{238}$U, $^{nat}$Pb and D$_2$O targets used in this work.
The $^{232}$Th target consisted of 5 identical disks of 2.00 mm thickness and 25.40 mm diameter. 
The disks are made of $^{nat}$Th which is 99.98\% $^{232}$Th and contains only trace amounts of $^{227-231, 234}$Th. 
The $^{238}$U target was composed of 8 disks of depleted uranium with thicknesses varying from $0.55-0.81$ mm. 
The $^{nat}$Pb target served as a `blank' to measure $\gamma$-ray beam induced backgrounds, and was machined to be
 comparable in areal density to the $^{232}$Th and $^{238}$U targets.
 A cylindrical D$_2$O cell made with an acrylic casing was used for characterizing the INVS 
 detector response to neutrons from the D($\gamma$,n) reaction.

\begin{table}[h]
\centering
\begin{tabular}{l | r | r | r}
\hline
Target & Mass (g) & Enrichment (\%) & Thickness (mm) \\ \hline
$^{232}$Th & $59.5\pm0.1$ & 99.98 & $10.00\pm0.02$\\ 
$^{238}$U & $53.67\pm0.01$ & $>99$& $5.06\pm0.03$\\
$^{nat}$Pb & $ 91.9\pm0.1 $ & $ - $& $ 10.6\pm0.1$\\
D$_{2}$O & $17.4\pm0.02$ & $>99.9$& $14.6\pm0.1$\\
\hline
\end{tabular}
\caption{Targets used in the experiment.}
\label{table:udisks}
\end{table}

\subsection{Neutron Detector}
\begin{figure}[h]
\includegraphics[width=3.375in]{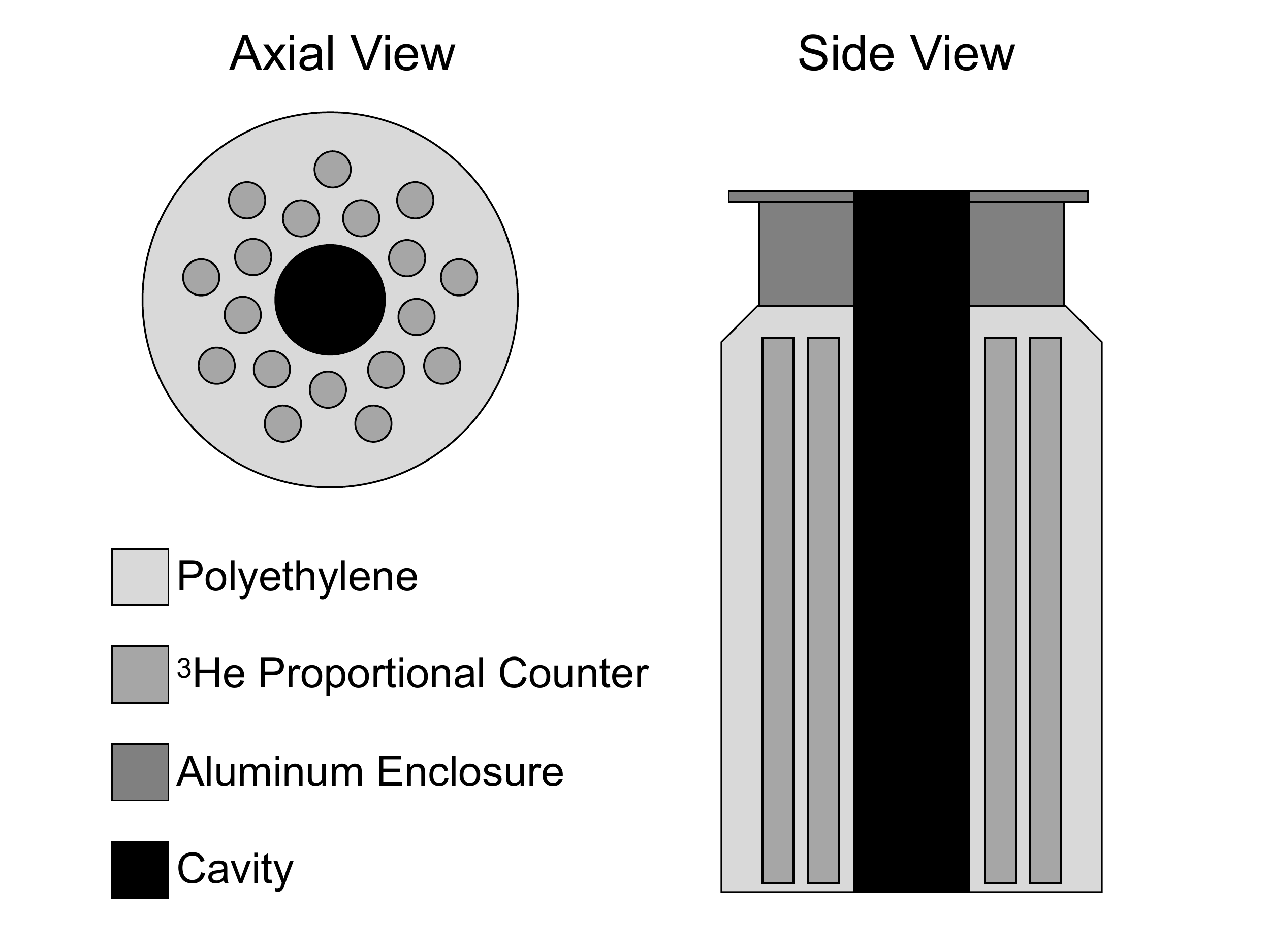}
\caption{Diagram of the INVS detector used in this work (not to scale). }\label{fig:invs_figure}
\end{figure}

Neutrons were detected with a model-IV Inventory Sample neutron detector \cite{sprinkle} (INVS), shown in
 Fig. \ref{fig:invs_figure}, consisting of 18 $^{3}$He proportional counters (PCs) embedded in a cylindrical shell of polyethylene 
moderator. The PCs had a diameter of 2.54 cm, an active length of 39 cm and a 
nominal $^3$He gas pressure of 6 atm. The PCs were arranged in 2 concentric rings with radii of 7.24 
cm and 10.60 cm, with each ring containing 9 equally spaced PCs. The polyethylene 
detector body was 46.2 cm long and 30.5 cm in diameter with an 8.9 cm diameter axial cavity for 
placing a neutron generating target.

As designed at Los Alamos National Laboratory, the INVS initially had preamplifier and discriminator 
circuits built into the detector. The INVS was modified for the present experiment to allow for single 
tube readout by installing SHV connections with direct access to each of the PC 
anodes. An array of Cremat CR-110 charge sensitive preamplifiers was mounted in a single enclosure 
with each preamplifier connected to a PC. The outputs of the preamplifiers 
were then sent to a CAEN V1730 500 MS/s 16-channel digitizer. The digitizer recorded 15 preamplifier 
channels, with the 16th channel reserved for a 1 Hz pulser for synchronizing digitizer time stamps and 
monitoring the DAQ live time.

\section{Detector Simulation \& Calibration}
\label{sec:simulation}

The full experimental setup including the INVS detector, targets, target holder assembly and detector 
shielding was modeled with a \textsc{geant4} Monte Carlo simulation  \cite{agostinelli2003}. The simulation 
was calibrated with efficiency measurements of the INVS detector made by Arnold
 \textit{et al.} \cite{ARNOLD201155}. The INVS neutron detection efficiency was measured on site using 
 the D($\gamma$,n) reaction, with statistical uncertainties of 1\% and systematic uncertainties of 
 $<3\%$. In order to fit the calibration data, the simulated INVS efficiency was scaled by 
 0.790$\pm$0.003 (stat) $\pm$0.02 (sys) for the inner ring and 0.775$\pm$0.004 (stat) $\pm$0.02 
 (sys) for the outer ring. These scaling factors account for loss of $^{3}$He gas pressure over time, 
 inefficiencies caused by the DAQ signal processing and discriminator threshold settings.

%The relative efficiency of each PC tube depends on a number of factors including the pressure of the 
%$^{3}$He gas, the threshold settings on the digitizer and the pulse shape discrimination (PSD) cuts. 
Since the $^{3}$He gas pressure, threshold settings and DAQ signal processing settings could be different
 for each PC, there was variation in the relative efficiency of each PC.
The relative PC tube efficiencies were measured by detecting neutrons from $^{232}$Th($\gamma$,f) 
reaction using circularly-polarized $\gamma$-ray beams, and were found to vary by $<10$\% from the 
average efficiencies of the inner and outer rings. 

The \textsc{geant4} simulation was validated through measurements of neutrons from the 
D($\gamma$,n) reaction using linearly-polarized, 4.3 MeV $\gamma$-ray beams. This reaction 
generates neutrons with well-defined energy and asymmetry in the $\phi$ angular distribution 
\cite{TORNOW20038,PhysRevC.61.061604}. The ratio of counts in the inner to outer ring of the INVS 
detector was measured to be 1.444$\pm$0.005, in agreement with the simulated ratio of 
1.46$\pm$0.05. The systematic uncertainty in the simulation comes from the uncertainties in the 
scaling factors for the inner and outer ring efficiencies. Enough events were included in the simulation 
to keep the statistical uncertainties more than an order of magnitude below that of the systematics. 

The angular distribution of neutrons from the D($\gamma$,n) reaction with linearly-polarized 
$\gamma$-ray beams has been explored in detail \cite{TORNOW20038} and a simple formalism was
developed to account for the polarization of the $\gamma$-ray beam \cite{PhysRevC.61.061604}. A 
D($\gamma$,n) reaction neutron generator was written and incorporated into the \textsc{geant4} 
simulation to compare the resulting asymmetries with the measured data. The simulation and 
measurement data were fit with a function of the form $a(1+b\cos2\phi)$, where $a$ is an overall 
scaling factor, $b$ is the detected asymmetry and $\phi$ is the angle of the PC relative to the polarization axis of 
the beam. The measurement yielded an asymmetry of 0.132$\pm$0.002 for the inner ring and 
0.252$\pm$0.003 for the outer ring, in agreement with the simulated asymmetries of 0.132$\pm$0.001 
and 0.248$\pm$0.002 for the inner and outer ring, respectively.

Since neutrons from a single fission event are highly correlated in angle and energy, accurately 
simulating the INVS detector efficiency required a source of event-by-event photofission neutrons.
The Monte Carlo-based code \textsc{freya} \cite{PhysRevC.80.024601} models fission observables in 
an event-by-event basis. By calculating neutron emission for specific instances of fission fragment 
masses and excitation energies, \textsc{freya} inherently provides correlations between the various 
observables, including the neutron energy and angle of emission relative to the direction of the fission 
fragments. \textsc{freya} only supports neutron-induced and spontaneous fission of several isotopes, 
so it was necessary to extend the code to model photofission of $^{232}$Th and $^{238}$U using the 
method described by Mueller \textit{et al.}  \cite{PhysRevC.89.034615}.

Fission product mass distributions were sampled from 0.5 MeV neutron induced fission data  
\cite{england7405evaluation} as suitable photofission data was not available. Fragment kinetic 
energies were also sampled from neutron induced fission data, with 2.97 MeV neutrons on $^{232}$Th  
\cite{HOLUBARSCH1971631} and 1.7 MeV neutrons on $^{238}$U \cite{VIVES200063}.

The \textsc{geant4} simulation sampled neutrons from data sets generated by \textsc{freya}, rotating 
the momentum of the neutrons so that the fission fragment axis followed the appropriate angular 
distribution. From the formalism in Refs.  \cite{RevModPhys.31.711, PhysRevC.85.014605}, fission 
fragments induced by a linearly polarized $\gamma$-ray beam have an angular distribution of the form
\begin{equation}
\begin{aligned}
W_{f}(\theta,\phi)=&a_{f}+b_{f}\sin^2(\theta)+c_{f}\sin^2(2\theta) \\
+&P_\gamma \cos(2\phi)\big( b_{f}\sin^2(\theta)+c_{f}\sin^2(2\theta)\big),
\end{aligned}
\end{equation}
where $\theta$ is the angle relative to the beam axis, $\phi$ is the azimuthal angle and $P_\gamma$ 
is the beam polarization. Assuming 100\% linear polarization \cite{Weller2009257} and neglecting the 
quadrupole contribution, the expression simplifies to
\begin{equation}
W_{f}(\theta,\phi)=a_{f}+b_{f}\sin^2(\theta)+b_{f}\cos(2\phi)\sin^2(\theta),
\label{eqn:fit}\end{equation}
where $a_{f}$ and $b_{f}$ are normalized such that $a_{f}+b_{f}=1$. Eq. \ref{eqn:fit} can also be used 
to fit the angular distribution of the neutrons emitted by the fragments, in order to give the correlation 
between the fragment distribution the neutron distribution. In this paper, the angular distribution is 
specified as being for the fission fragments or fission neutrons by the use of the subscripts $_{f}$ and 
$_{n}$, respectively.

\section{Data Reduction \& Analysis}
\subsection{Pulse Shape Discrimination}

The digitized waveforms of the preamplifier outputs of the PCs were analyzed offline. When a trigger 
occurred in any channel the digitizer recorded a 4.1 $\mu$s long, 2050 sample trace for all 16 
channels. In addition each trigger had an associated timestamp based on a 125 MHz internal clock in 
the digitizer.  A 1 Hz pulser in one of the digitizer channels served as the DAQ live time monitor. The 
maximum trigger rate that the digitizer could handle without data loss was approximately 1 kHz, so 
caution was taken to keep the trigger rate below  $\sim500$ Hz by limiting the $\gamma$-ray beam 
flux with attenuators at higher energies where the photofission cross sections were greater. As long as 
the trigger rate was kept reasonable, the 1 Hz pulser consistently indicated no dead time.

$^3$He-based PCs are relatively insensitive to $\gamma$-rays but the large $\gamma$-ray beam flux 
of $\sim10^8$ $\gamma$/s provided by HI$\gamma$S combined with a thick actinide target created a 
detectable background of Compton scattered $\gamma$-rays. The scattered $\gamma$-rays interact 
with the PCs primarily by scattering electrons into the active gas volume. These 
$\gamma$-ray induced events have the potential to produce pulses with a sufficiently large amplitude 
that they overlap with the pulse height spectrum from neutron induced events, in which a proton and 
triton created by the $^3$He(n,p)$^3$H reaction deposit a maximum of 763.8 keV into the active gas volume of the 
PC.

Identification of neutron and $\gamma$-ray detection events was achieved through the use of a pulse shape discrimination (PSD) 
technique originally developed for $^{3}$He proportional counters in low background experiments 
\cite{LANGFORD201351}. This technique exploits the difference in stopping power between the  
ions in a neutron detection event and the electron in a $\gamma$-ray detection event. For the same 
deposited energy the proton and triton will have a shorter track length than the electron, and 
thus a shorter pulse rise time.

\begin{figure}[h]
\centering
\includegraphics[width=3.375in]{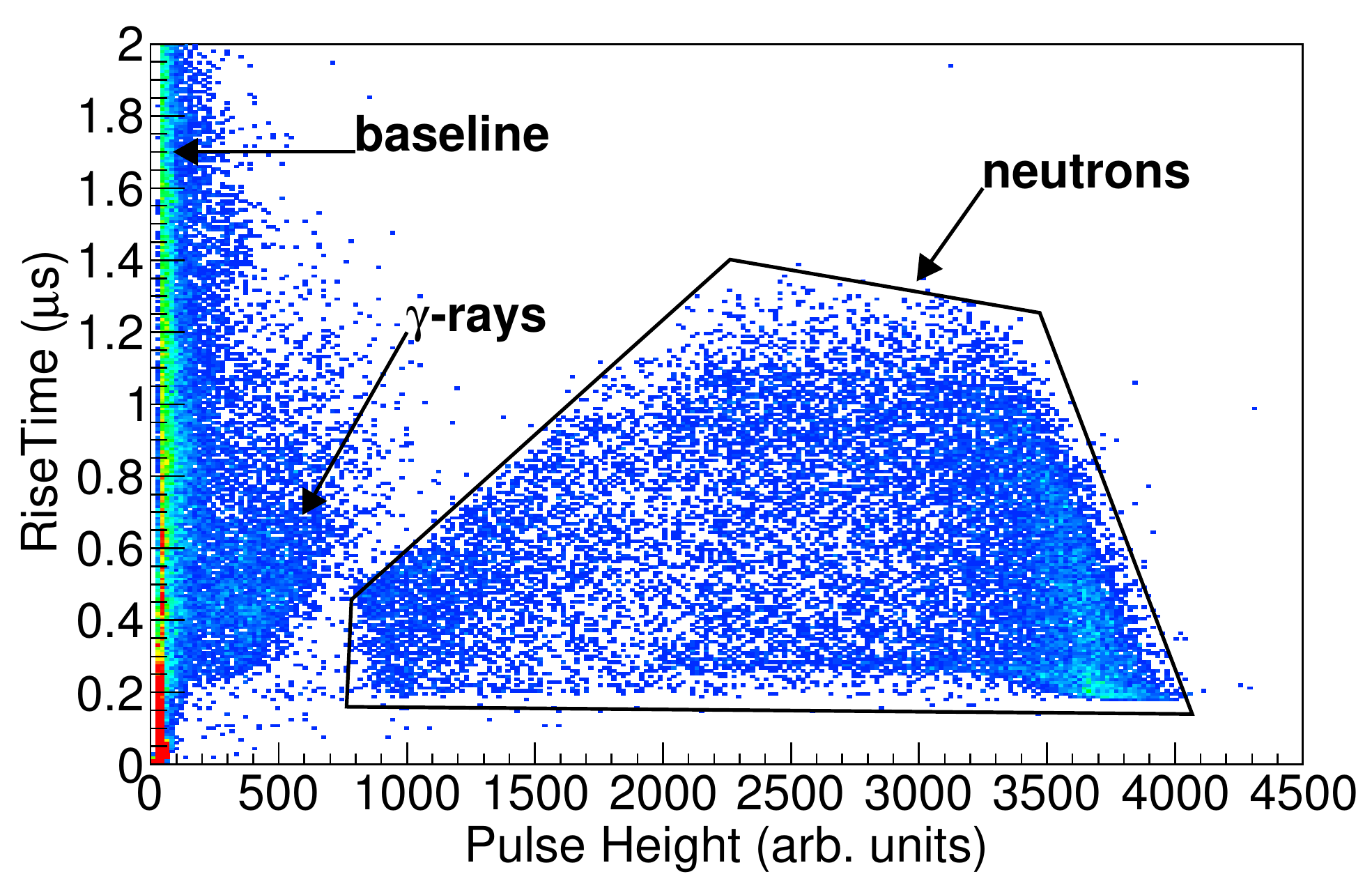}
\caption{(Color online) PSD plot from a PC tube for a $\gamma$-ray beam incident on a $^{232}$Th target, with a neutron cut region.}\label{fig:psd_plot}
\end{figure}

Following the application of a Gaussian smoothing routine to reduce high-frequency noise, the rise 
time, pulse height and timestamp of each digitized waveform were extracted. The rise time was 
defined as the time for a pulse to rise from 10\% to 50\% of its maximum pulse height above the 
baseline. Fig. \ref{fig:psd_plot} shows the PSD cut used for one of the $^3$He proportional counters. 
The neutron and $\gamma$-ray event regions are well separated in rise time vs pulse height space. 
Measurements made with a $^{nat}$Pb target for generating Compton scattered $\gamma$-rays and a 
deuterium target for generating neutrons confirmed the particle identification regions in the PSD plot. Events in the
baseline noise region occur because the analysis code attempts to extract pulse characteristics from 
each digitizer channel, regardless of which channel triggered the DAQ since coincident neutron 
detections happened regularly. In the absence of a neutron or $\gamma$-ray event waveform, the 
analysis routine extracts pulse characteristics from random noise.

\subsection{Background Neutron Multiplicity Analysis}
\label{sec:bgmult}

%Need to add time distribution info

Determining the detected neutron multiplicities requires setting a time window in which neutrons from 
the same fission event may be counted. The time window must be long enough that any correlated 
neutrons from a single fission event will have either been detected or escaped the detector; however, 
making it excessively long increases the likelihood that uncorrelated neutrons from other fission events 
or backgrounds are counted as well. Thus a clear understanding of the neutron detection timescale is 
necessary. 

\begin{figure}[h]
\centering
\includegraphics[width=3.375in]{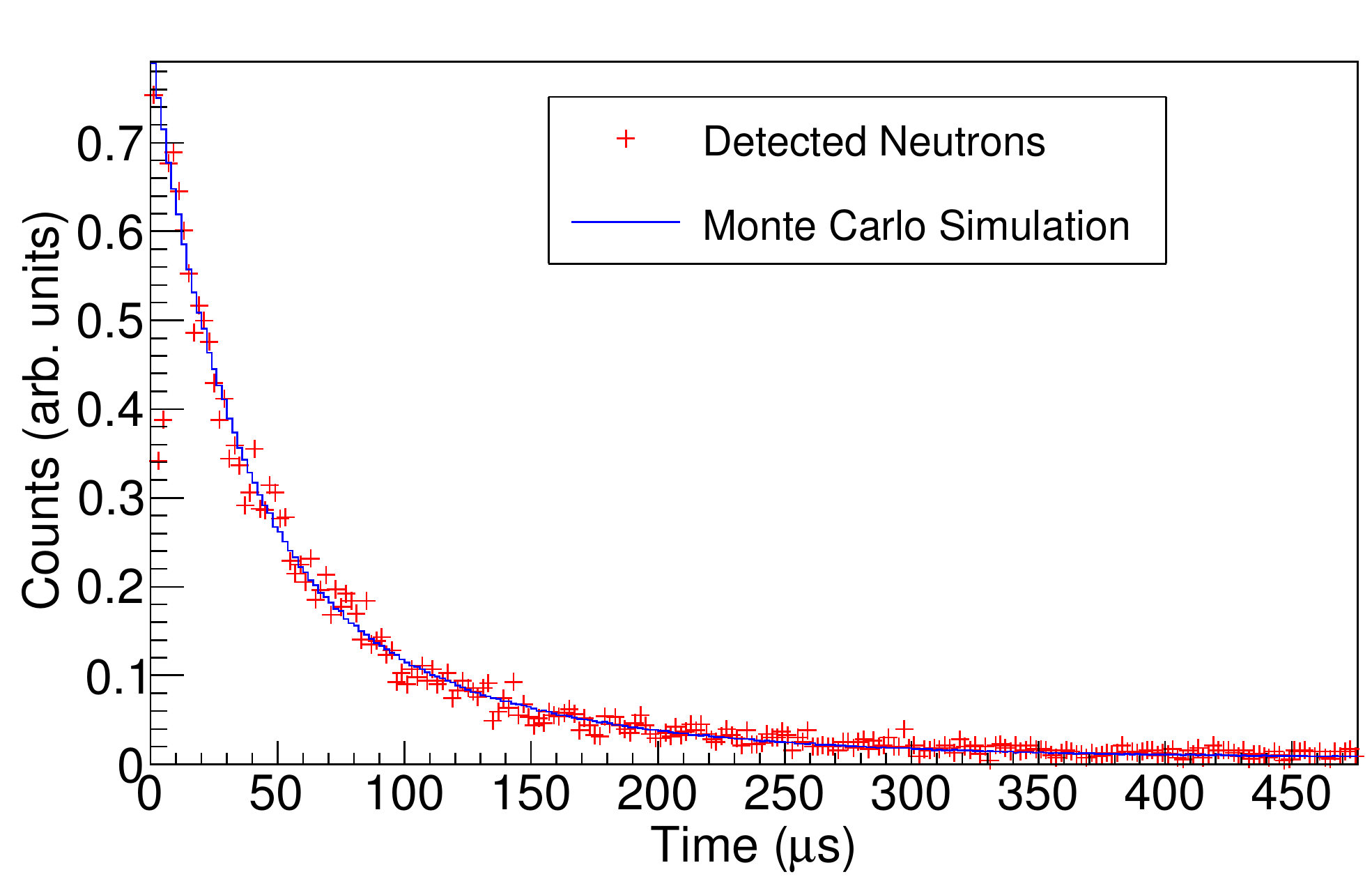}
\caption{(Color online) Comparison of experimental data and a Monte Carlo simulation of neutron time distributions in 
the INVS detector for the $^{238}$U($\gamma$,f) reaction at E$_\gamma$=5.1 MeV.}\label{fig:time_distribution}
\end{figure}

\textsc{geant4} simulations of the INVS detector were performed and the detection time for each 
neutron was recorded, where the neutron is always emitted at time $t=0$ and detected some time after 
that. The simulated neutron detection time distribution was found to be well represented by an 
exponential decay function with a 31 $\mu s$ half life fit. A direct comparison of neutron detection time 
cannot be generated from the experimental data because there is no `start' signal for each fission 
event. The closest comparison is to instead sort through the detected neutrons in chronological order, 
with the first detected neutron creating a gate and subsequent neutron detection times recorded 
relative to the first one. A 1 ms long gate was used to guarantee that all detected neutrons from the 
fission event were included. The analysis was applied to Monte Carlo simulation data and 
experimental data from the $^{238}$U($\gamma$,f) reaction at E$_\gamma$=5.1 MeV, an energy at which the 
photofission rate was much larger than the background event rate but still sufficiently low that 
coincident fission events were rare. Fig. \ref{fig:time_distribution} shows the good agreement between
 the time distributions for simulated and detected photofission neutrons in the INVS detector, validating the 
neutron detection time response of the \textsc{geant4} simulation.

\begin{figure}[h]
\centering
\includegraphics[width=3.375in]{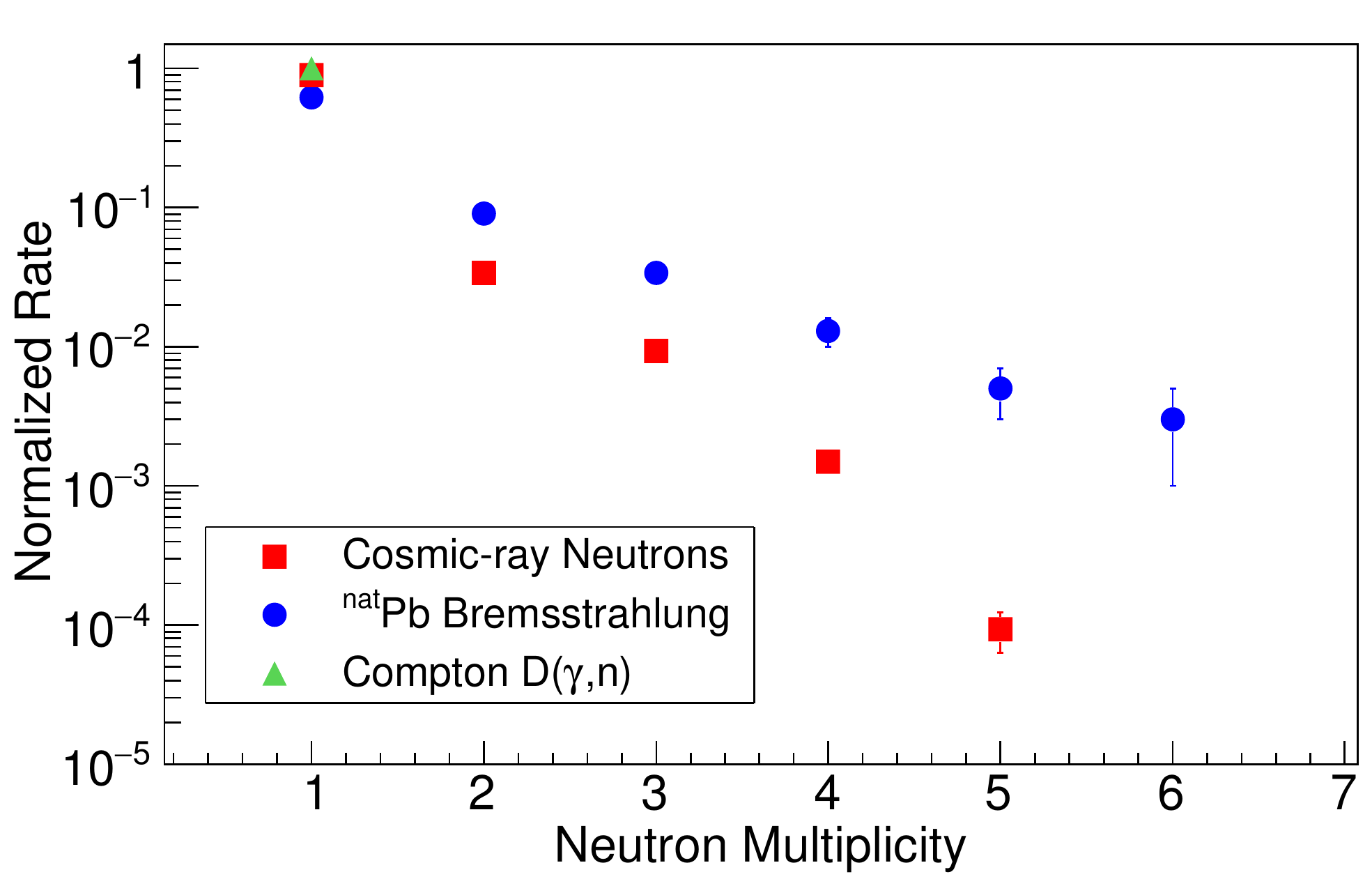}
\caption[(Color online) Multiplicity distributions for various backgrounds]{Neutron multiplicity rates, normalized to 1 
Hz, from backgrounds caused by cosmic-ray induced neutrons, bremsstrahlung contamination of the 
$\gamma$-ray beam inducing $^{nat}$Pb($\gamma$,xn) reaction neutrons, and Compton scattered 
$\gamma$-rays inducing D($\gamma$,n) reaction neutrons in the polyethylene moderator of the INVS 
detector.}\label{fig:bgmultcomp}
\end{figure}

There were three sources of background neutrons present throughout this work: cosmic-ray induced 
neutrons, neutrons from the D($\gamma$,n) reaction occurring in the detector moderator, and 
neutrons from bremsstrahlung contamination in the HI$\gamma$S beam \cite{SCOTTCARMAN19961} inducing ($\gamma$,xn) and ($\gamma$,f)
reactions on the targets. Sources of background neutrons were differentiated with a multiplicity 
analysis technique. Background multiplicities were defined by grouping together neutrons within 300 
$\mu$s coincidence windows, where the first detected neutron defines the start of the coincidence 
gate. The multiplicity was defined as the total number of neutrons in the gate including the one which 
triggered it, meaning that the minimum detected multiplicity was 1 by definition. The next 300 $\mu$s 
long gate was created on the first neutron that fell outside of the previous gate so that any neutron was 
only counted towards one multiplicity event. The observed neutron multiplicity distributions for the 
various backgrounds are shown in Fig. \ref{fig:bgmultcomp}.

The cosmic-ray induced neutron background was reduced to a neutron detection rate of 
$0.537\pm0.003$ Hz by surrounding the INVS detector with $\sim30$ cm of borated polyethylene 
shielding on all sides.

A bremsstrahlung component of the $\gamma$-ray beam was identified by placing a $^{nat}$Pb target 
in the INVS detector and operating HI$\gamma$S in ``single bunch mode," in which a single electron 
bunch was held in the FEL storage ring instead of the two used in normal operation 
\cite{Weller2009257}. In this configuration there are no counter-propagating electrons so the 
HI$\gamma$S facility is incapable of generating $\gamma$-rays by Compton scattering the FEL 
photons. Further, the configuration of the HI$\gamma$S facility was such that $\gamma$-rays 
generated by Compton scattered FEL photons would have been below the threshold for 
$^{nat}$Pb($\gamma$,xn) reactions. 
Determining the properties of the bremsstrahlung beam is beyond the scope of this work.
We sought to only measure the directly observable effects of the bremsstrahlung: the neutron detection rates and detected neutron multiplicity distribution.
The observed bremsstrahlung-induced $^{nat}$Pb($\gamma$,xn) neutron multiplicity distribution is shown in Fig. \ref{fig:bgmultcomp},
with a typical neutron detection rate of $\sim 0.75-1.5$ Hz.

\begin{figure}[h]
\centering
\includegraphics[width=3.375in]{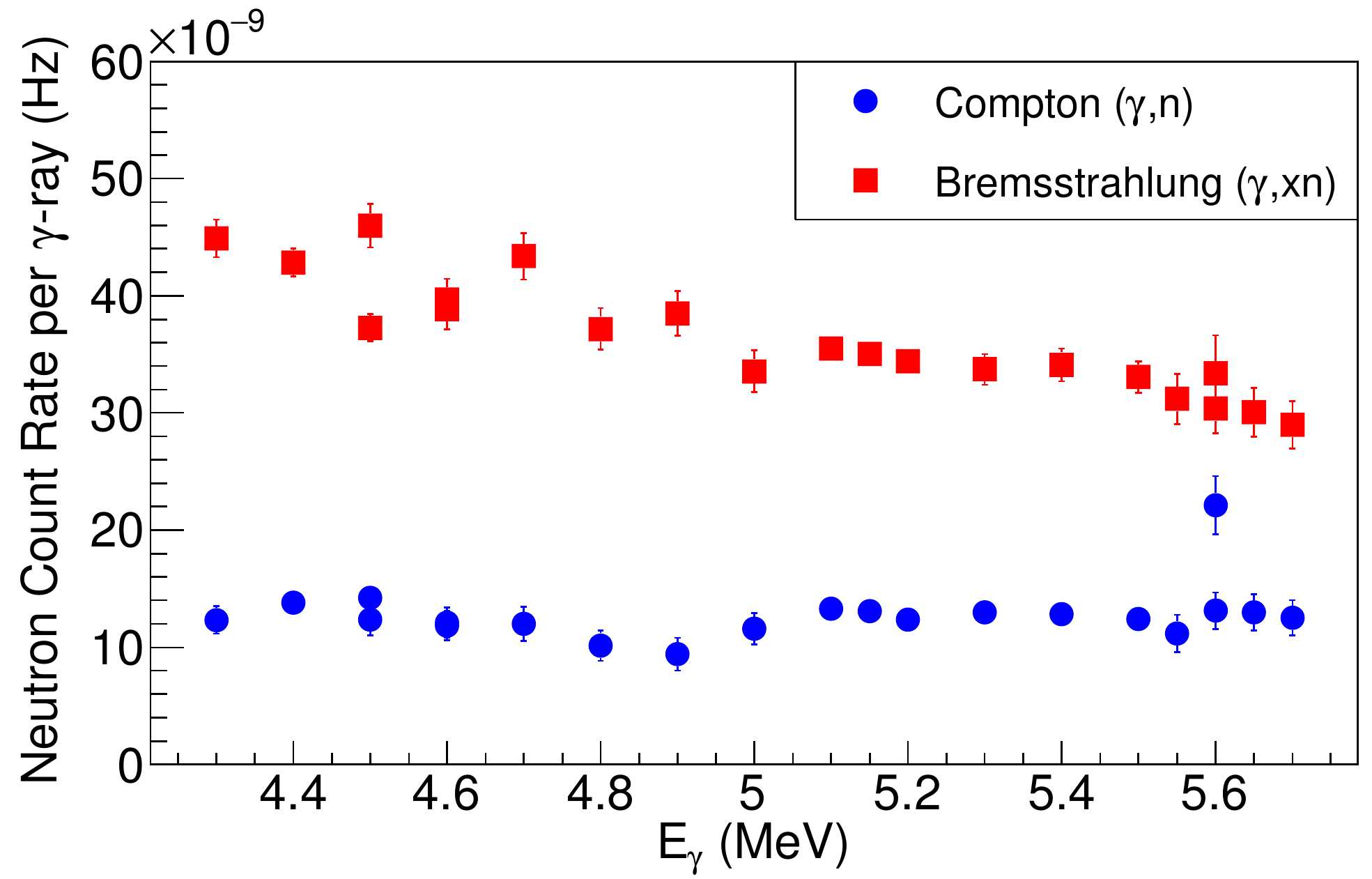}
\caption[(Color online) Measured HI$\gamma$S beam-induced backgrounds]{HI$\gamma$S beam-induced 
background neutron count rates observed with a $^{nat}$Pb target, normalized to the flux of the 
primary $\gamma$-ray beam.}\label{fig:higs_background_rate}
\end{figure}

The combined bremsstrahlung and Compton scattering induced backgrounds were measured by 
placing a $^{nat}$Pb target in the INVS detector and running HI$\gamma$S in the standard ``two bunch 
mode" which produces an intense, quasi-monoenergetic $\gamma$-ray beam. The observed neutron 
multiplicity distributions were fit with neutrons from the $^{nat}$Pb($\gamma$,xn) reaction and 
Compton scattering D($\gamma$,n) reaction multiplicity distributions to determine the contributions from 
the two backgrounds, with typical uncertainties of $\leq10\%$. The background neutron detection rates 
as a function of E$_\gamma$ are shown relative to the primary $\gamma$-ray beam flux in Fig. 
\ref{fig:higs_background_rate}. The Compton scattering D($\gamma$,n) reaction and bremsstrahlung 
($\gamma$,xn) reaction backgrounds were both scaled appropriately for the $^{232}$Th and 
$^{238}$U targets. The Compton scattering component scales with the number of electrons in the 
target nuclei, the target thickness and density. The bremsstrahlung component was multiplied by 
scaling factors for each target which were experimentally measured using the ``single bunch mode" 
HI$\gamma$S operation.

\subsection{Prompt Fission Neutron Multiplicity Analysis}\label{sec:mult_analysis}
Defining the detected neutron multiplicities by grouping together neutrons within a 300 $\mu$s window 
works well for the low event rates in the background analysis (see Sec. \ref{sec:bgmult}), but suffers 
from pileup effects for the higher event rates of photofission measurements. Consequently the prompt 
fission neutron analysis relied on the Rossi-alpha method  \cite{pacilio1969reactor}, a multiplicity logic 
scheme originally developed for reactor neutron noise analysis. This method is more robust against 
accidental coincidences caused by high fission rates.

The gating logic used in the present work was as follows: after a neutron detection, there was a 10 
$\mu$s delay to account for a small dead time immediately following the digitizer gate. Next a 500 
$\mu$s long gate was set which contained real and accidental coincidences (RA), and following that a 
500 $\mu$s long gate which should only contain accidental coincidences (A). Each neutron detection 
created the RA and A gates, and the number of events of each multiplicity was given by RA-A, the 
difference between the real events plus accidental events and just the accidentals.

Even with the Rossi-alpha multiplicity gating logic, the detected neutron multiplicity distributions 
depend on the fission event rate and overlap with the background in a nonlinear manner. A Monte 
Carlo simulation was developed to fit the prompt fission neutron multiplicity distribution with the 
detected Rossi-alpha multiplicities. The fission neutron multiplicity distribution was modeled as a 
Gaussian distribution \cite{PhysRev.108.783} with the fit parameters mean, $\overline{\nu}$, and 
spread, $\sigma$. The Monte Carlo simulation modeled the detection of the photofission neutrons with 
the assumed initial distribution, added the background events with the experimentally measured 
multiplicities and rates (see Sec. \ref{sec:bgmult}), and analyzed the resulting data stream using the 
Rossi-alpha gating logic. The multiplicity distribution parameters $\overline{\nu}$ and $\sigma$ were 
varied to minimize the $\chi^2$ between the measured and the Monte Carlo simulated RA-A 
distributions.

\subsection{Neutron Asymmetry Analysis}

\begin{figure}[!htb]
\includegraphics[width=3.375in]{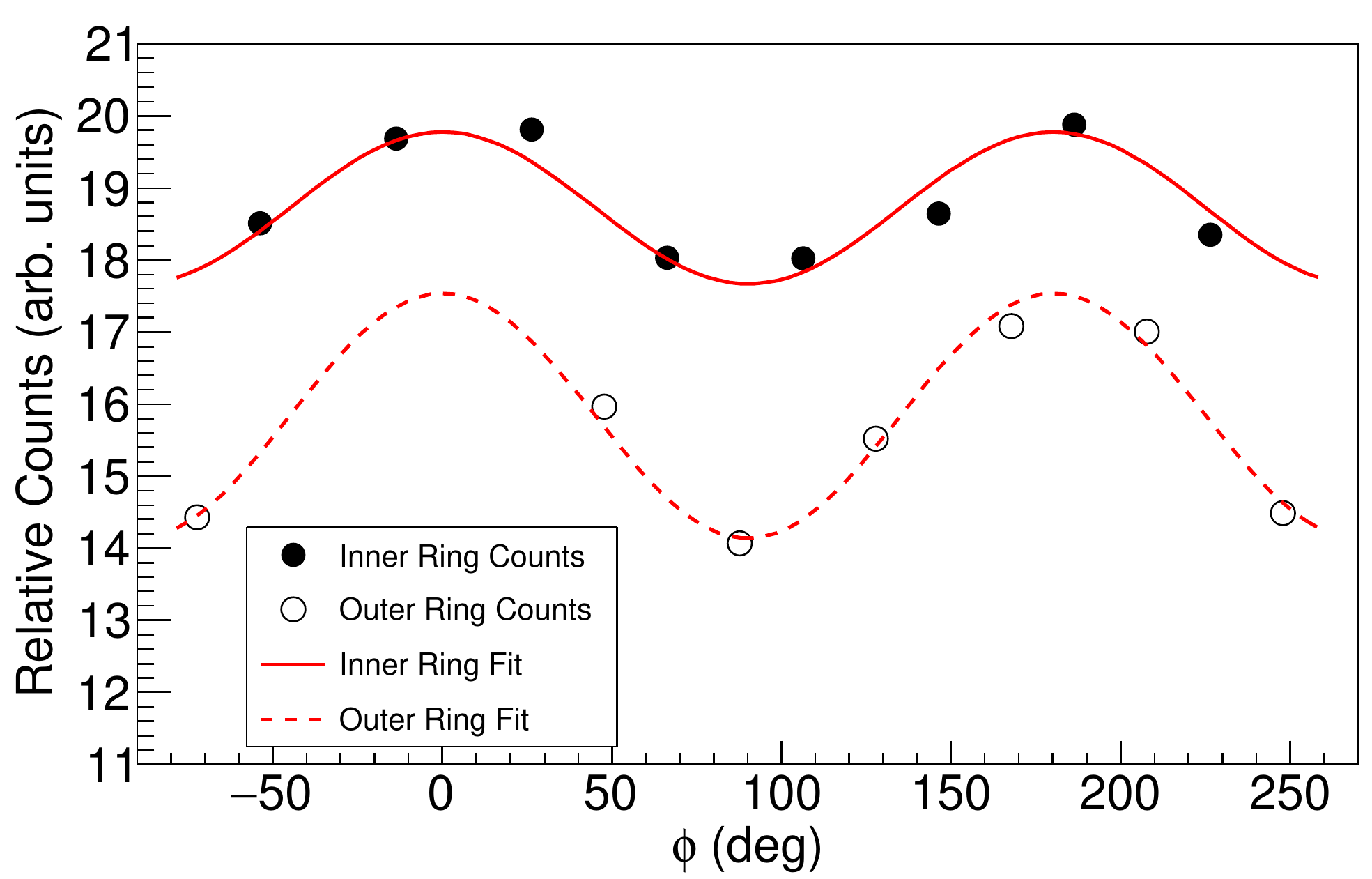}
\caption{(Color online) Detected neutron yields and asymmetry in the inner and outer rings of PC tubes in the INVS 
detector for E$_\gamma$  = 5.6 MeV on a $^{238}$U target. The outer ring counts have been multiplied by a 
factor of 1.3 to appear on the same scale as the inner ring. Statistical error bars are smaller than the markers.}\label{fig:uranium_asymmetry_fit}
\end{figure}

After subtracting background counts and correcting for the relative PC efficiencies, the detected 
prompt photofission neutron asymmetries were measured by fitting the relative yields of each PC tube 
with the function 
\begin{equation}
Y_{d}(\phi) = a_{d}\big(1+b_{d}\cos(2\phi)\big),
\label{eqn:fit_asyms}
\end{equation}
where $Y_{d}(\phi)$ is the detected neutron yield in the PC tube at angle $\phi$, $a_{d}$ is an overall scaling 
factor and $b_{d}$ is the detector asymmetry. Fig. \ref{fig:uranium_asymmetry_fit} shows examples of 
the fit, performed independently on the inner and outer rings of the INVS detector.

\begin{figure}[h]
\centering
\includegraphics[width=3.375in]{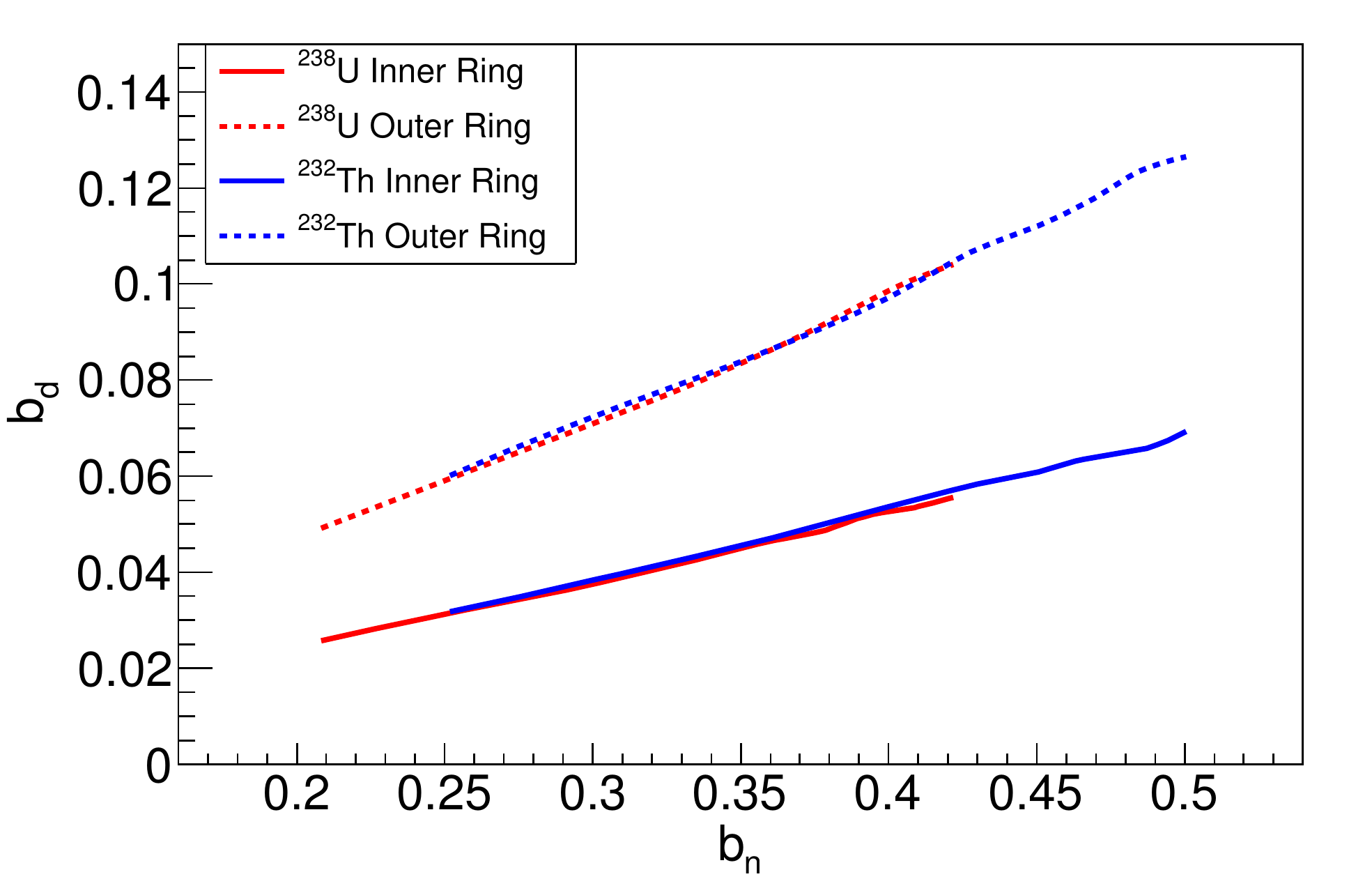}
\caption{(Color online) Simulated correlation between the detected asymmetry $b_{d}$ and emitted neutron 
polarization asymmetry $b_{n}$ for the $^{238}$U and $^{232}$Th photofission neutrons.}\label{fig:bd_bn}
\end{figure}

The simulated correlation between the detected neutron asymmetry $b_d$ and the emitted neutron 
asymmetry $b_n$ is shown in Fig. \ref{fig:bd_bn}, where $b_n$ is defined in Eq. \ref{eqn:fit}. 
\textsc{geant4} simulations were performed using the photofission neutrons from the \textsc{freya} calculations for 
both $^{238}$U and $^{232}$Th, with each unique target geometry modeled to account for neutron 
scattering in the target and target holder assemblies. Since the inner and outer rings of the INVS 
detector have different detected asymmetry responses to the same emitted neutron angular 
distribution, each ring was treated as a separate measurement of the polarization asymmetry and the 
results were combined for a single measurement for each E$_\gamma$ and target. There was no 
strong dependence observed between the fissioning isotope species and the detector response to the 
neutron asymmetry, which is consistent with the results of Ref. \cite{PhysRevC.89.034615}.

\subsection{Photofission Cross Section Analysis}

The photofission cross section is written as
\begin{equation}
\sigma(\gamma,f)=\frac{(N_n - N_b)A_t}{\epsilon_{INVS}\,  \overline{\nu}\,  N_\gamma \, \ell_t \, \rho_t  
f N_A},
\end{equation}
where $N_n$ is the total number of detected neutrons, $N_b$ is the number of background neutrons, 
$A_t$ is the atomic mass of the target nuclei, $\epsilon_{INVS}$ is the efficiency of the INVS detector, 
$\overline{\nu}$ is the mean prompt neutron multiplicity, $N_\gamma$ is the total number of 
$\gamma$-rays on target, $\ell_t$ is the target thickness, $\rho_t$ is the target density, $f$ is a factor 
which accounts for the attenuation of the $\gamma$-ray beam within the thick target and $N_A$ is 
Avogadro's number. The thick target correction factor is written as 
\begin{equation}
f=\frac{1-e^{-\mu  \rho_t  \ell_t}}{\mu \, \rho_t \, \ell_t},
\end{equation}
where $\mu$ is the attenuation coefficient of the target material \cite{nistxray}. $N_b$ is determined 
using the multiplicity analysis technique described in Sec. \ref{sec:mult_analysis}.

The photofission neutron spectra from the \textsc{freya} calculations did not differ enough over the 
range of 4.3 MeV $<$ E$_\gamma<6.0$ MeV to change the simulated efficiency of the INVS detector. 
Thus $\epsilon_{INVS}=0.295\pm0.009$ for the $^{232}$Th($\gamma$,f) neutrons and 
$\epsilon_{INVS}=0.277\pm0.008$ for the $^{238}$U($\gamma$,f) neutrons. The values of 
$\overline{\nu}$ for $^{232}$Th($\gamma$,f) and $^{238}$U($\gamma$f) were determined by taking the 
weighted average of the multiplicity measurements results for each target (see Sec. \ref{sec:mult}). 
The photofission cross section analysis in this work uses $\overline{\nu}_{mean}$ = 
$2.22\pm^{0.02}_{0.05}$ and $\overline{\nu}_{mean}$ = $2.46\pm^{0.01}_{0.03}$ for $^{232}$Th and 
$^{238}$U, respectively. The $\sim3\%$ systematic error in the photofission cross section data includes contributions from $\epsilon_{INVS}$ (3\%),
 target thickness (0.2\% for $^{232}$Th, 0.6\% for $^{238}$U, 1\% for $^{nat}$Pb and 0.7\% for D$_2$O) and mean neutron multiplicities (0.4-2\%).

\section{Results \& Discussion}

\subsection{Prompt Photofission Neutron Polarization Asymmetries}

\begin{figure}[!htb]
\centering
\includegraphics[width=3.375in]{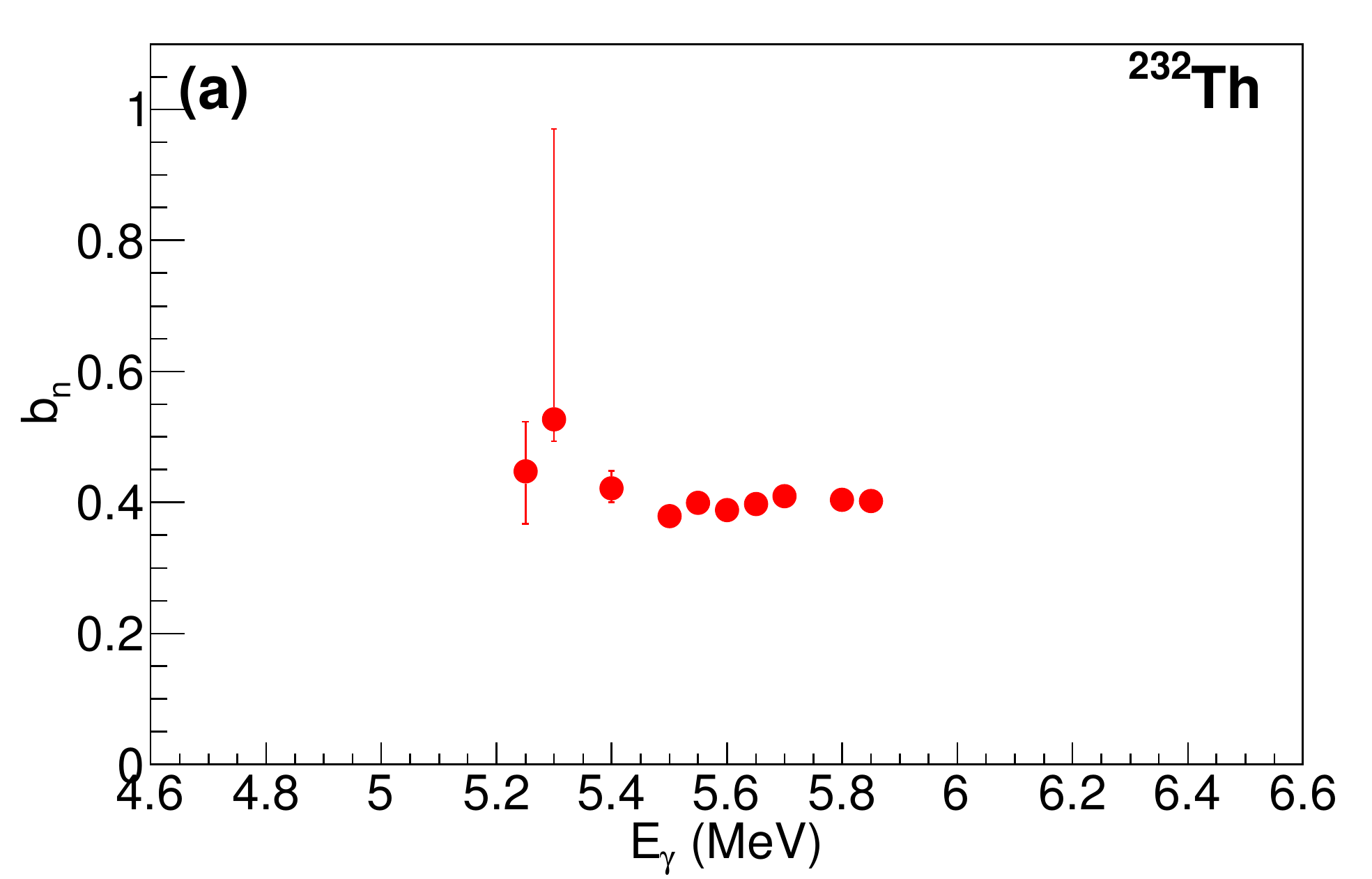}
\includegraphics[width=3.375in]{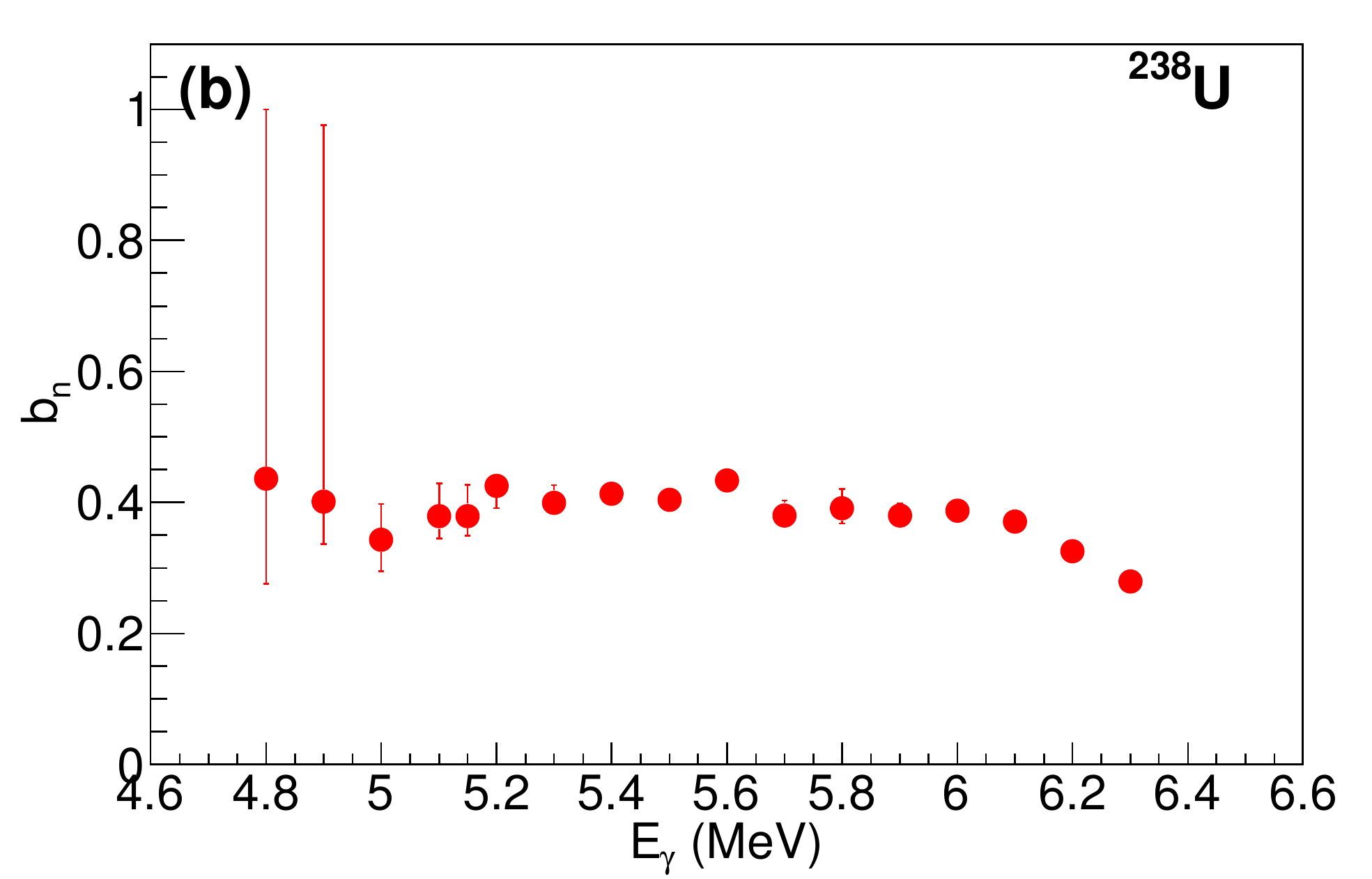}
\caption{(Color online) Photofission neutron polarization asymmetries for (a) $^{232}$Th and (b) $^{238}$U. 
The error bars reflect the fit error from Eq. \ref{eqn:fit_asyms}.}\label{fig:asymmetry_results}
\end{figure}

The measured prompt photofission polarization asymmetries are shown in Fig. 
\ref{fig:asymmetry_results} for $^{238}$U and $^{232}$Th, and are qualitatively consistent with the 
expected effects of the transmission through the fission barrier. Based on the the most likely energy 
level ordering for an even-even nucleus \cite{HUIZENGA1967614}, the lowest energy $J^\pi = 1^-$ 
excitation is $(J^\pi , K) = (1^-, 0)$ which corresponds to the mass asymmetry mode. This would then 
be the dominant fission channel at low  E$_\gamma$ and would result in large polarization 
asymmetries. The next lowest channel would be the $(1^-, \pm1)$ bending mode which would begin to 
contribute as E$_\gamma$  is increased, reducing the polarization asymmetry, as experimentally 
observed.

\begin{figure}[!htb]
\centering
\includegraphics[width=3.375in]{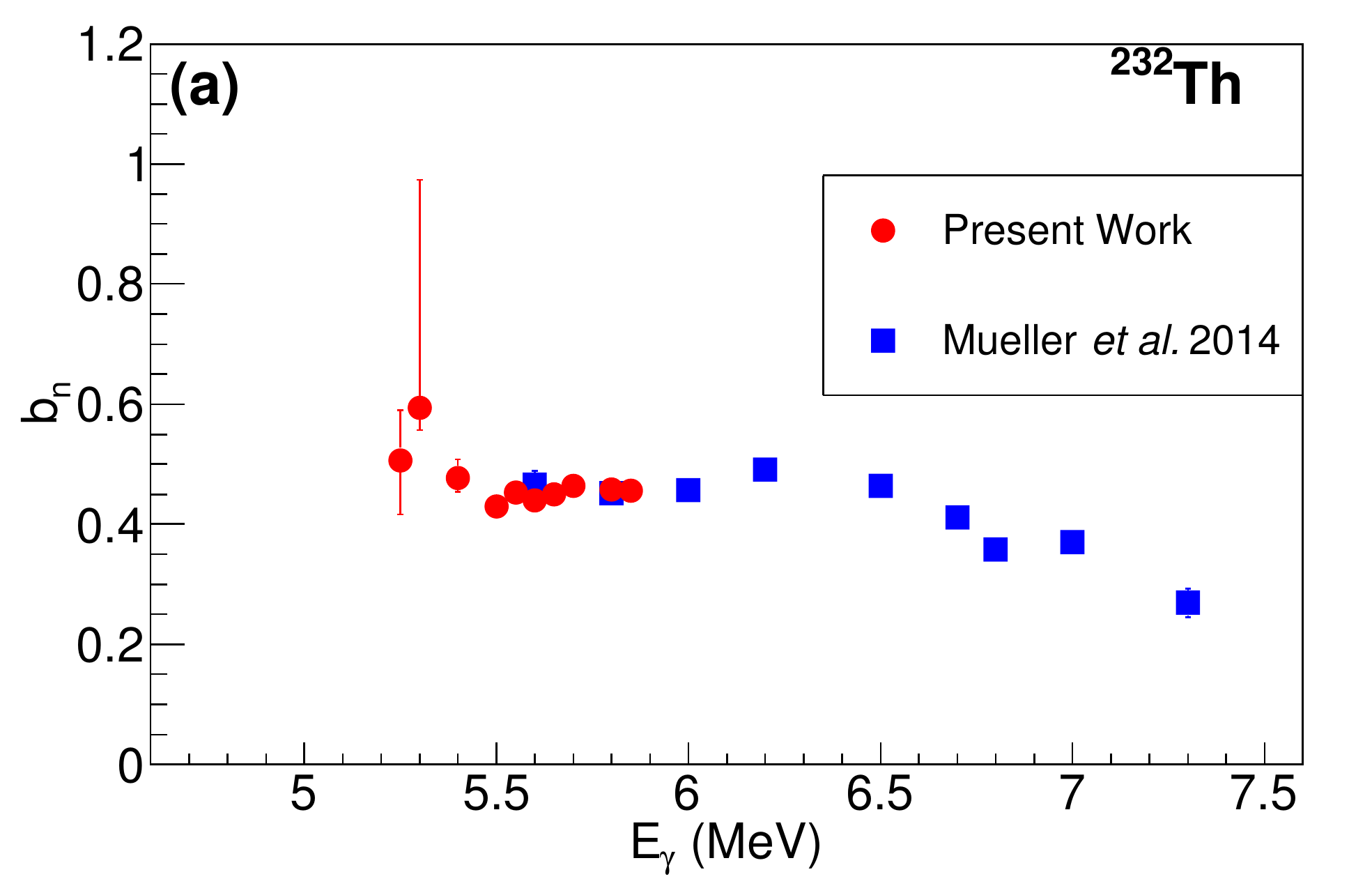}
\includegraphics[width=3.375in]{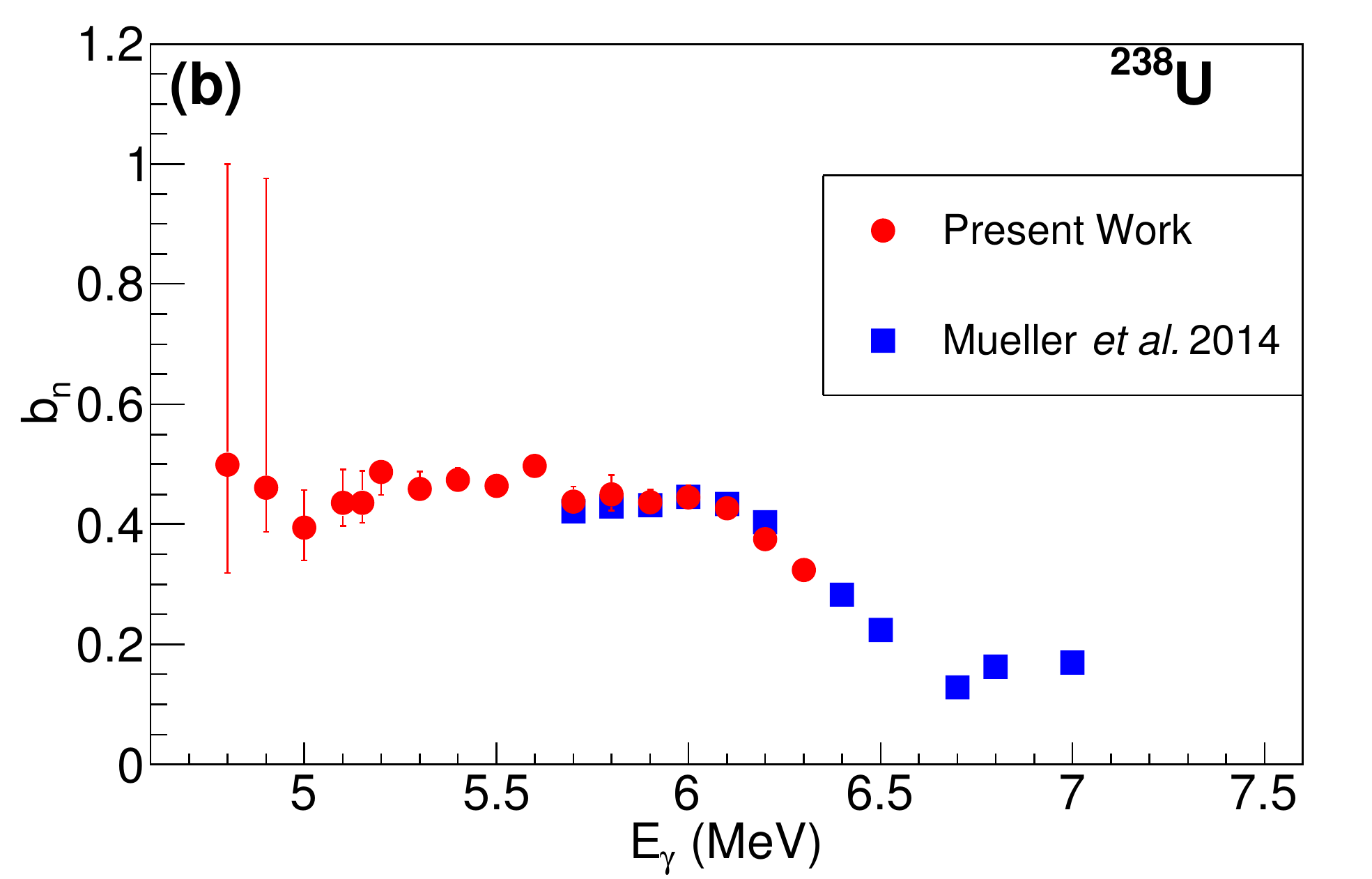}
\caption{(Color online) Photofission neutron polarization asymmetries in (a) $^{232}$Th and (b) $^{238}$U for 
neutrons with E$_n>$1.5 MeV, compared with the data of Mueller \textit{et al.} 
\cite{PhysRevC.89.034615}.}\label{fig:jon_comparison}
\end{figure}

In Fig. \ref{fig:jon_comparison} the present asymmetry results are compared with the data of Mueller 
\textit{et al.} \cite{PhysRevC.89.034615}, the only other measurement of photofission neutron 
polarization asymmetries in $^{238}$U and $^{232}$Th. Since liquid scintillating neutron detectors 
were used in Ref. \cite{PhysRevC.89.034615}, the presented data were limited to neutrons above E$_n=1.5$ 
MeV. In order to make a direct comparison with the prior data, the present asymmetry calculations were adjusted 
to include only neutrons above E$_n= 1.5$ MeV. This adjustment was achieved by setting a 1.5 MeV 
energy threshold on the neutron distribution fit which correlates the fission fragment asymmetry with 
the emitted neutron asymmetry. Omitting the lower energy neutrons increases $b_n$ by about 10\% 
since the neutrons which are emitted in the direction of the fragment receive more of a kinematic 
boost, and therefore tend to have higher energies than those emitted perpendicular to the fragment 
motion. Once the effects of the neutron energy cut of E$_n>1.5$ MeV are accounted for, there is excellent agreement between the present results and the data of Mueller \textit{et al.} in the energy region where they overlap.

\subsection{Prompt Photofission Neutron Multiplicities}\label{sec:mult}

\begin{figure}[h]
\centering
\includegraphics[width=3.375in]{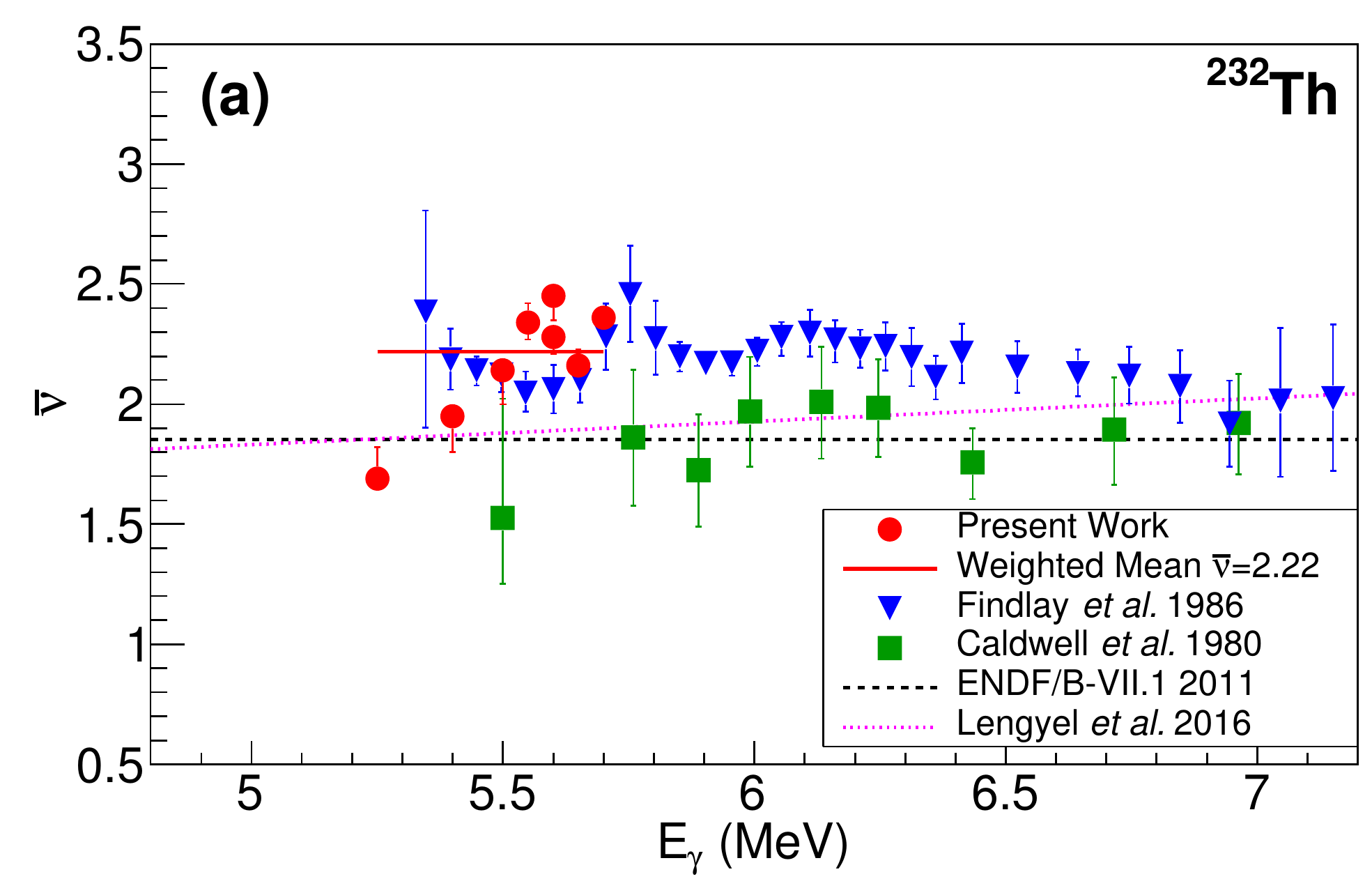}
\includegraphics[width=3.375in]{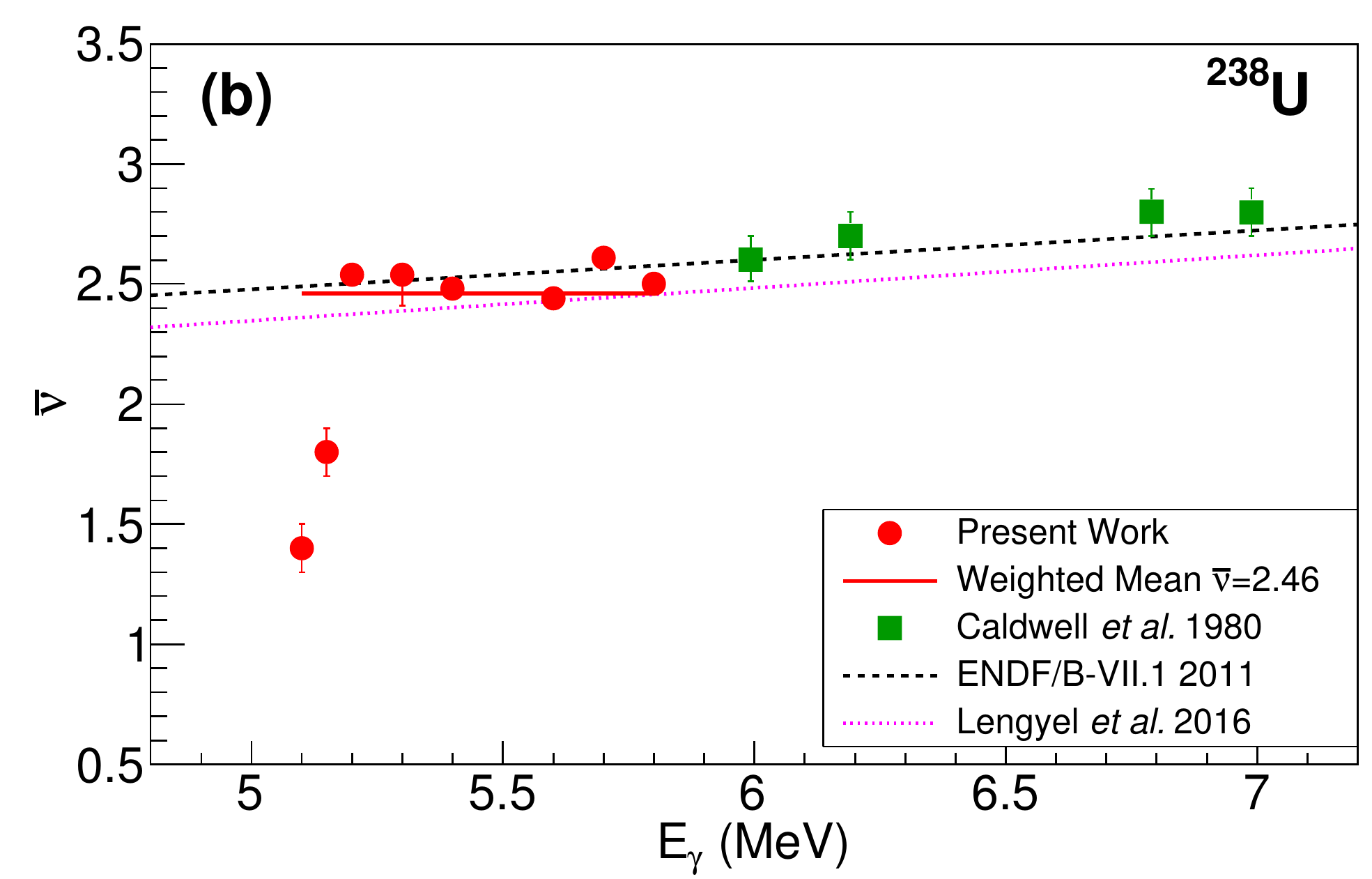}
\caption{(Color online) Measured mean photofission neutron multiplicities for (a) $^{232}$Th and (b) $^{238}$U, 
compared with the data of Caldwell \textit{et al.} \cite{Caldwell_1980}, Findlay \textit{et al.} 
\cite{FINDLAY1986217}, the ENDF/B-VII.1 evaluation \cite{CHADWICK20112887}, and the empirical 
model of Lengyel \textit{et al.} \cite{lengyel}.}\label{fig:mean_mults}
\end{figure}

The mean prompt photofission neutron multiplicities, $\overline{\nu}$, for $^{232}$Th and $^{238}$U 
are shown in Fig. \ref{fig:mean_mults}, along with previous measurements \cite{Caldwell_1980, 
FINDLAY1986217}, the ENDF/B-VII.1 evaluation \cite{CHADWICK20112887}, and an empirical model 
developed by Lengyel \textit{et al.} \cite{lengyel}. The error bars in the present data represent the 
range of values that increase the $\chi^2$ of the fit by less than 1. A weighted mean of the present 
data was calculated using the inverse of the $\chi^2$ as the weighting factor. Multiplicities determined 
in the present work are $\overline{\nu}_{mean}$ = $2.22\pm^{0.02}_{0.05}$ and 
$\overline{\nu}_{mean}$ = $2.46\pm^{0.01}_{0.03}$ for the photofission of $^{232}$Th and $^{238}$U, 
respectively. The lowest E$_\gamma$  points which diverge from the rest of the data suffer from poor 
statistics and consequently have larger $\chi^2$ values exceeding 200 and 50 for the lowest 
E$_\gamma$ measurements for $^{232}$Th  and $^{238}$U, respectively. In the case of $^{232}$Th, 
the present data are in better agreement with Findlay \textit{et al.} \cite{FINDLAY1986217} than with the data of Caldwell 
\textit{et al.} \cite{Caldwell_1980}. The measurements of Ref. \cite{Caldwell_1980} appear systematically low in comparison, along with the ENDF/B-VII.1 evaluation \cite{CHADWICK20112887} and the empirical model \cite{lengyel}.
In the case of $^{238}$U, the present data are in good agreement with the data of Ref. \cite{Caldwell_1980}, the ENDF/B-VII.1 evaluation \cite{CHADWICK20112887} and the empirical model of Ref. \cite{lengyel}.

\begin{figure}[h]
\centering
\includegraphics[width=3.375in]{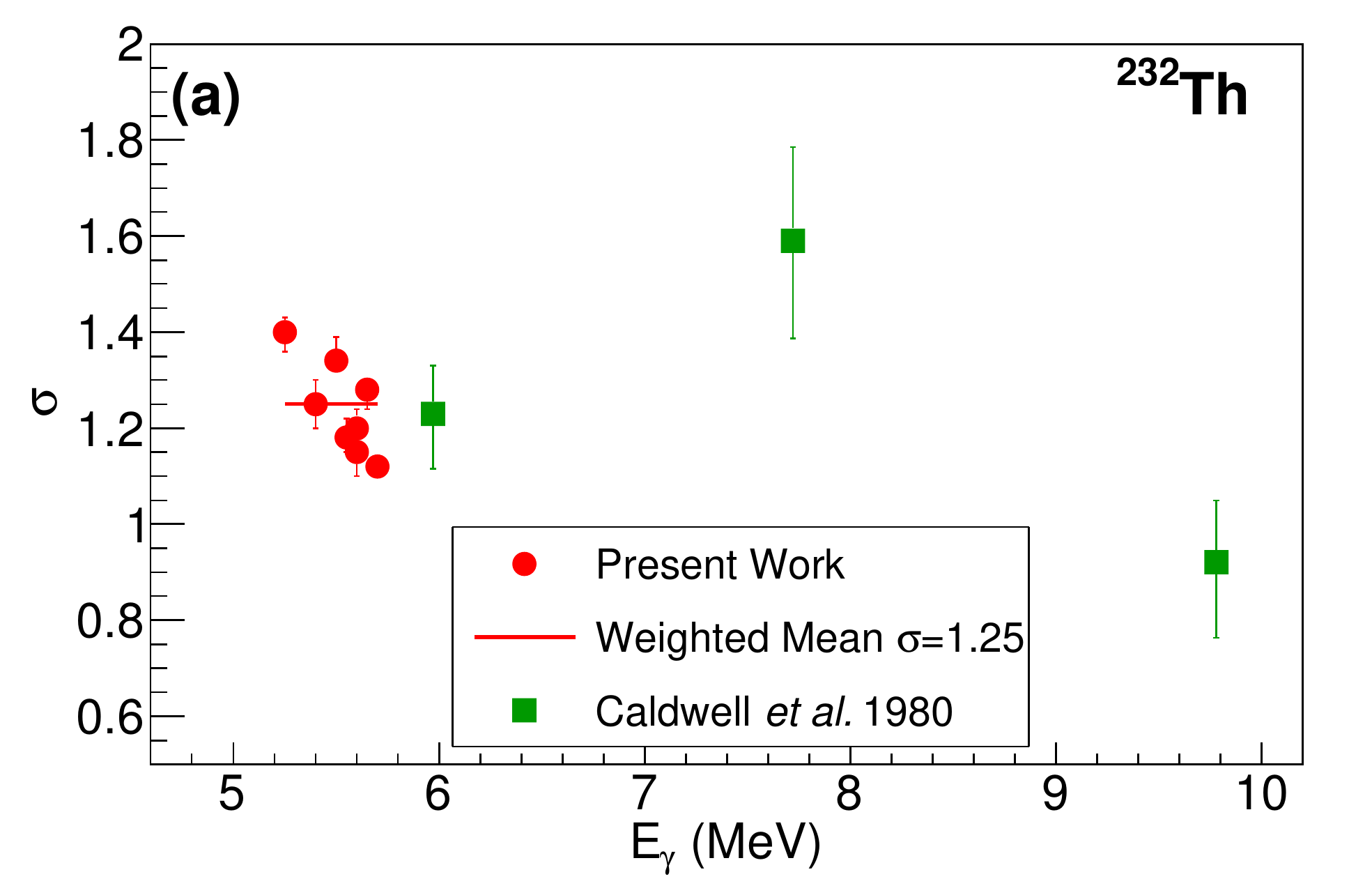}
\includegraphics[width=3.375in]{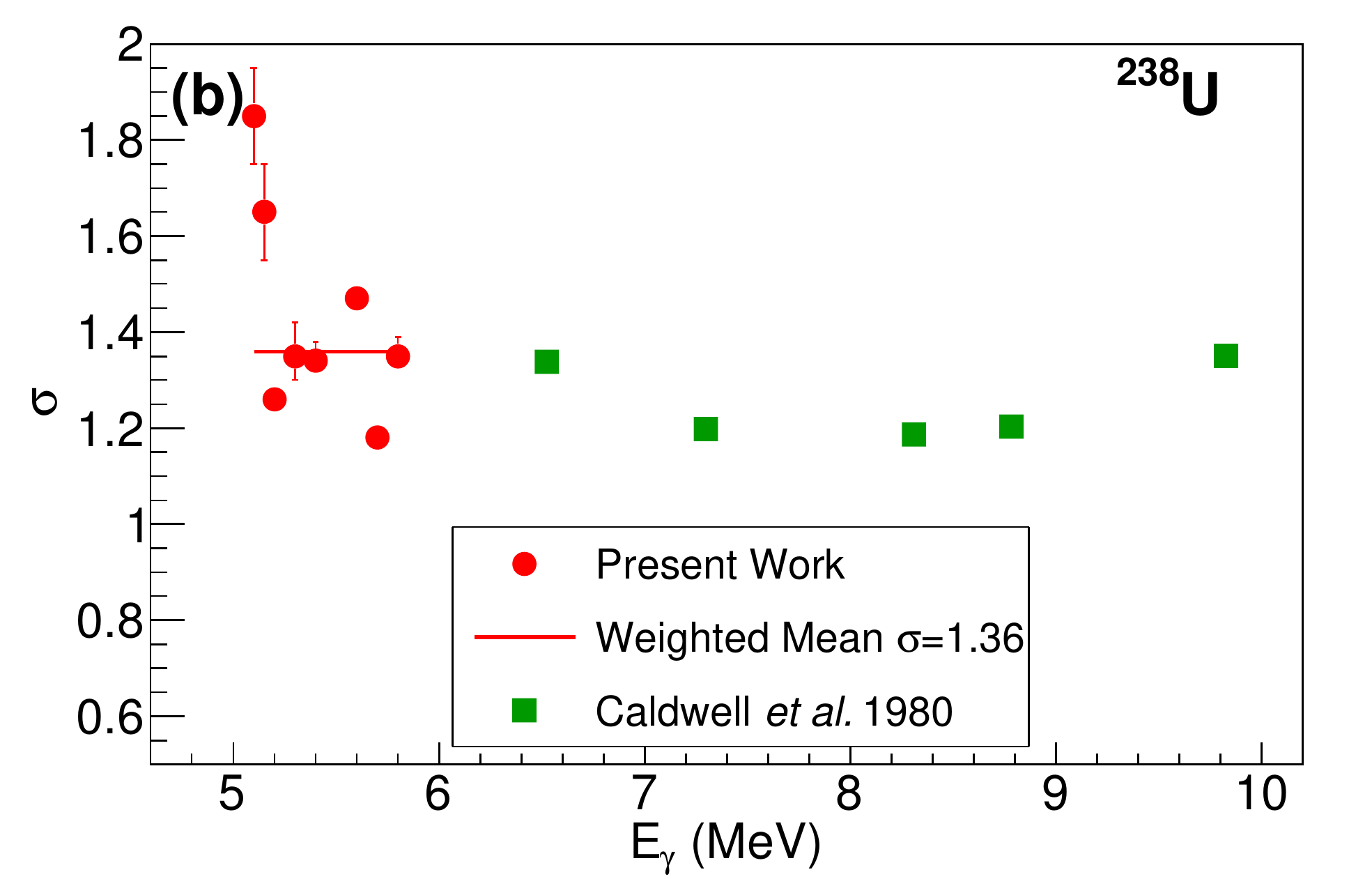}
\caption{(Color online) Measured spreads of the prompt neutron multiplicity distributions for (a) $^{232}$Th and (b) 
$^{238}$U, compared with the data of Caldwell \textit{et al.} \cite{Caldwell_1980}.}\label{fig:sigma_mults}
\end{figure}

The spreads of the Gaussian prompt neutron multiplicity distributions, $\sigma$, are plotted in Fig. 
\ref{fig:sigma_mults}, in good agreement with the only previous measurement in a comparable 
E$_\gamma$ range \cite{Caldwell_1980}. The $\chi^2$ weighted means were measured to be 
$\sigma_{mean}=1.25\pm^{0.02}_{0.01}$ and $\sigma_{mean}=1.36\pm^{0.02}_{0.01}$ for 
$^{232}$Th  and $^{238}$U, respectively.

\subsection{Photofission Cross Sections}

\begin{figure*}[!htb]
\centering
\includegraphics[width=6.75in]{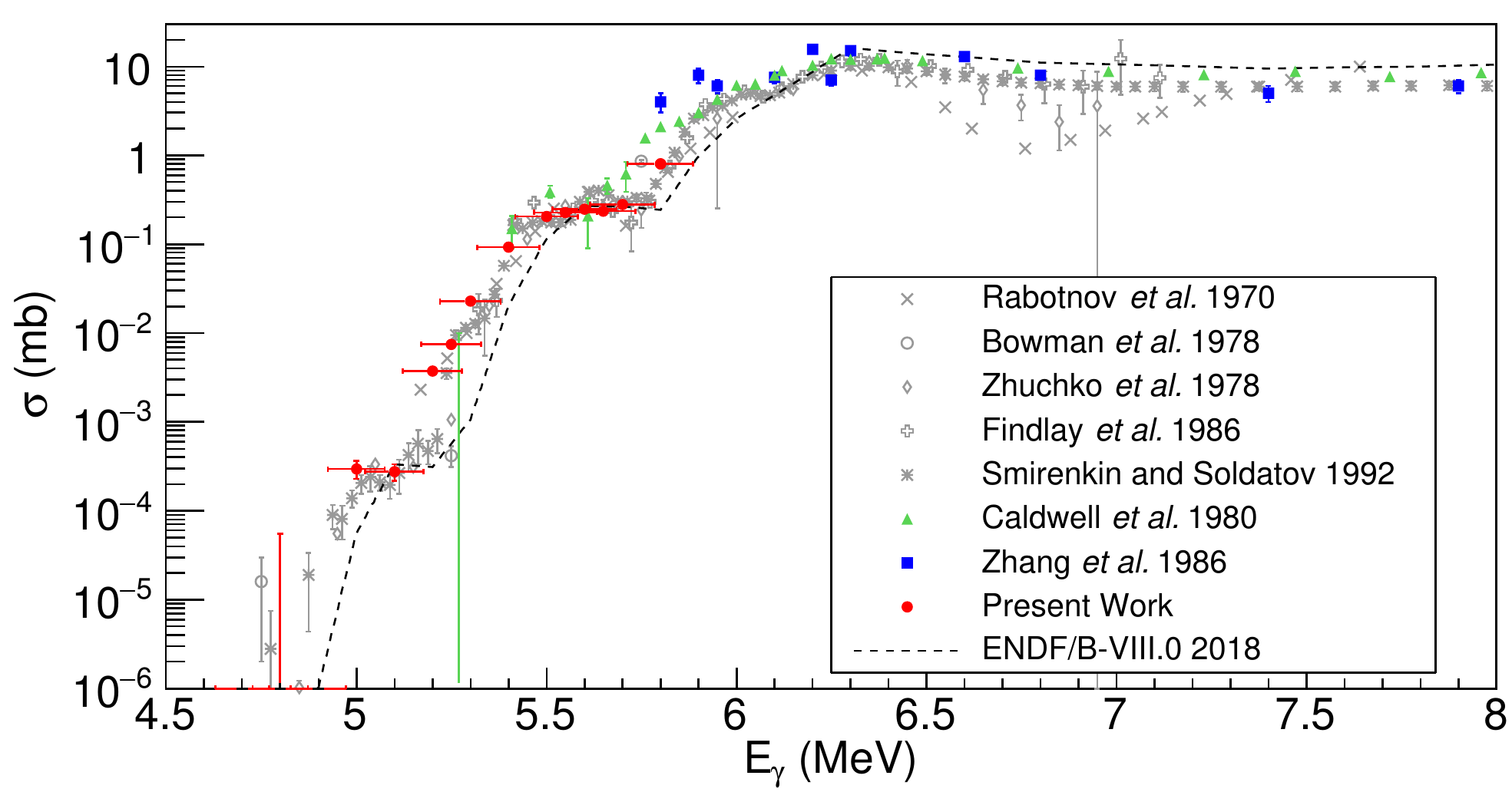}
\caption{(Color online) The measured photofission cross section of $^{232}$Th (red circles) compared with data obtained using  
bremsstrahlung $\gamma$-ray beams (various gray markers) 
\cite{rabotnov1970photofission, PhysRevC.17.1086, zhuchko1978study, FINDLAY1986217, Smirenkin:1996aa}, data obtained using
quasi-monoenergetic $\gamma$-ray beams (green triangles, blue squares)
 \cite{PhysRevC.21.1215, PhysRevC.34.1397} and the ENDF/B-VIII.0 evaluation (dashed line) 
 \cite{BROWN20181}. Vertical error bars represent statistical uncertainty and horizontal bars represent 
 the energy resolution of the $\gamma$-ray beams. }\label{fig:thorium_contrast_cross_sections}  
\end{figure*}

The $^{232}$Th($\gamma$,f) reaction cross section results are listed in Table \ref{table:thfission} and 
plotted in Fig.\ref{fig:thorium_contrast_cross_sections}, along with all available literature data sets which 
extend below an excitation energy of 6 MeV  
\cite{rabotnov1970photofission, PhysRevC.17.1086, zhuchko1978study, FINDLAY1986217, Smirenkin:1996aa, PhysRevC.21.1215, PhysRevC.34.1397} 
and the ENDF/B-VIII.0 evaluation \cite{BROWN20181}. Data sets consistent with 0 mb below $\sim6$ 
MeV  
\cite{KNOWLES1982315, PhysRevLett.35.501, YESTER1973593, MAFRA1972110, KHAN1972333} 
were omitted.
\begin{table}[h]
\centering
\begin{tabular}{ c  c  c }
\hline
\hline
E$_\gamma$ (MeV) \vline& $\sigma_{\text{E}_\gamma}$ (MeV) \vline& $\sigma(\gamma$,f) ($\mu$b) \\ \hline
4.7 & 0.071 & -0.4 $\pm$ 0.2\\
4.8 & 0.072 & -0.04 $\pm$ 0.1\\
4.9 & 0.074 & -0.4 $\pm$ 0.1\\
5.0 & 0.075 & 0.30 $\pm$ 0.07\\
5.1 & 0.077 & 0.27 $\pm$ 0.06\\
5.2 & 0.078 & 3.73 $\pm$ 0.09\\
5.25 & 0.079 & 7.4 $\pm$ 0.2\\
5.3 & 0.080 & 22.7 $\pm$ 0.2\\
5.4 & 0.081 & 92.9 $\pm$ 0.3\\
5.5 & 0.083 & 203.8 $\pm$ 0.5\\
5.55 & 0.083 & 227.3 $\pm$ 0.5\\
5.6 & 0.084 & 248.0 $\pm$ 0.5\\
5.65 & 0.085 & 235.1 $\pm$ 0.5\\
5.7 & 0.089 & 277.1 $\pm$ 0.5\\
5.8 & 0.087 & 799 $\pm$ 3\\
 \hline
\end{tabular}
\caption{Tabulated $^{232}$Th($\gamma$,f) reaction cross section data. The measurements were 
performed with Gaussian $\gamma$-ray beam spectra with mean E$_\gamma$ and spread 
$\sigma_{\text{E}_\gamma}$. The quoted errors are the statistical uncertainties, and there is an 
additional overall 3\% systematic error}
\label{table:thfission}
\end{table}

There is some tension between the present data and the ENDF/B-VIII.0 evaluation, with the present 
data about an order of magnitude higher in the 5.2-5.5 MeV range. In this region the cross section in the ENDF/B-VIII.0 
evaluation is lower than nearly all of the available experimental data.
The present results are generally in good agreement with the data obtained using bremsstrahlung beams 
\cite{rabotnov1970photofission, PhysRevC.17.1086, zhuchko1978study, FINDLAY1986217, Smirenkin:1996aa}.
 In particular the present work observes the same plateau in the photofission cross section of 
 $^{232}$Th in the E$_\gamma$  range of $ 5.4-5.7$ MeV. Blokhin and Soldatov \cite{Blokhin2009} 
 attribute this plateau to an almost complete fragmentation of a resonance in the second minimum 
 caused by damping, and a partial fragmentation of a resonance in the third minimum which is shifted 
 in energy relative to the second minimum resonance. This combination of resonant states in the 
 second and third minima explains the large width of the plateau and the presence of resonance 
 structure. However, the 5.6 MeV resonance observed in the data of Smirenkin and Soldatov \cite{Smirenkin:1996aa} is not 
 seen in the present results. This resonance is also not apparent in the only other data measured with 
 quasi-monoenergetic $\gamma$-ray beams in that energy region \cite{PhysRevC.21.1215}.
 Despite the absence of the 5.6 MeV resonance in present results, the $ 5.4-5.7$ MeV plateau that is apparent in all 
 available data sets is difficult to explain without damped vibrational states, which is suggestive of a deep third minimum in the fission barrier.

\begin{figure*}[!htb]
\centering
\includegraphics[width=6.75in]{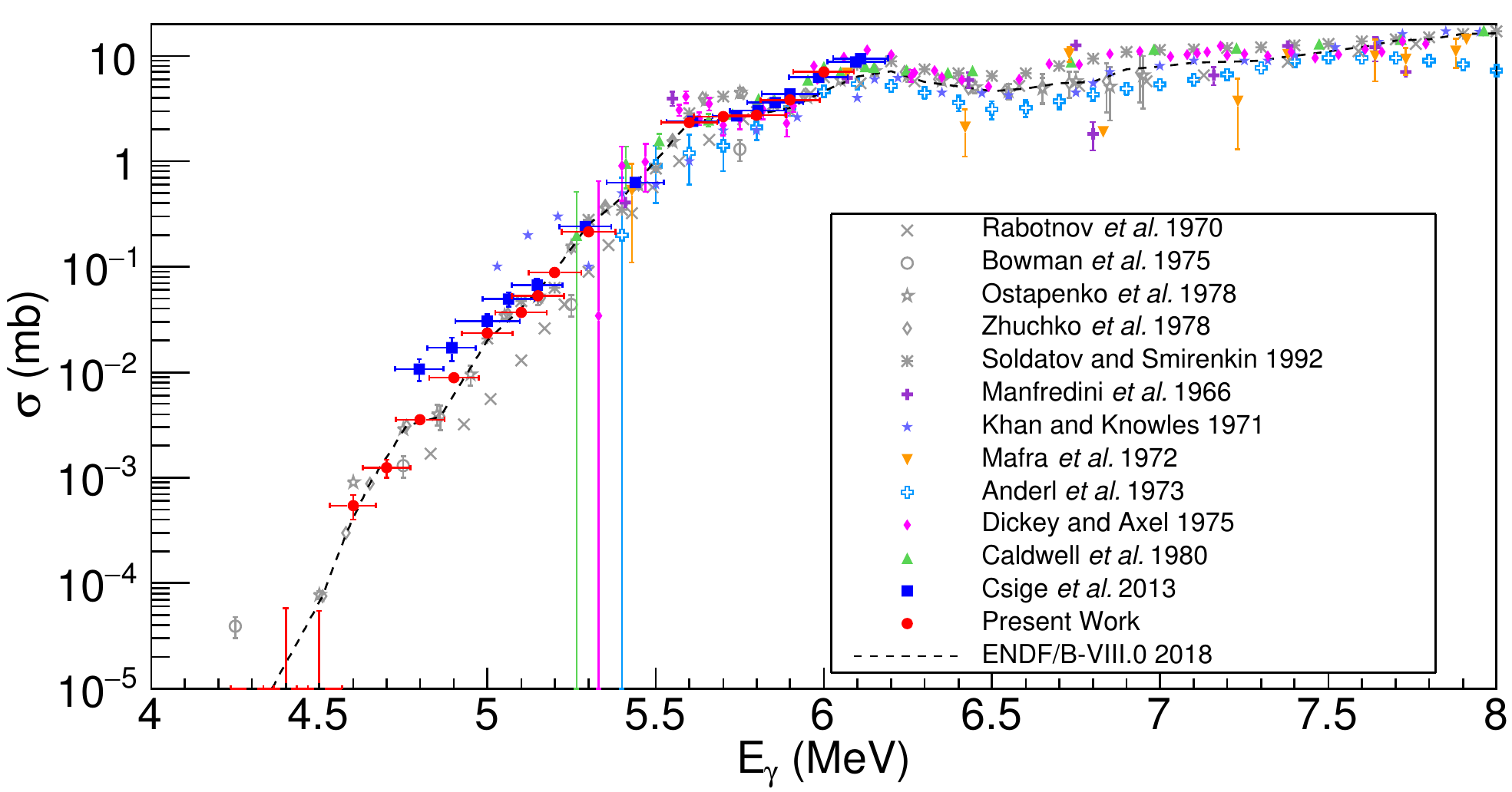}
\caption{(Color online) Measured photofission cross section of $^{238}$U (red circles) compared 
with data obtained using bremsstrahlung $\gamma$-ray beams (various gray markers)
 \cite{rabotnov1970photofission, bowman863, ostapenko1978yields, zhuchko1978study, soldatov1992yield}, 
 data obtained using quasi-monoenergetic $\gamma$-ray beams (purple crosses, blue stars, orange triangles, blue crosses, pink diamonds, 
 green triangles, blue squares) 
 \cite{manfredini1966results, KHAN1972333, MAFRA1972110, ANDERL1973221, PhysRevLett.35.501, PhysRevC.21.1215, PhysRevC.87.044321} 
 and the ENDF/B-VIII.0 evaluation (dashed line) \cite{BROWN20181}. Vertical error bars represent statistical uncertainty and horizontal bars 
 represent the energy resolution of the $\gamma$-ray beams. }\label{fig:uranium_contrast_cross_sections}  
\end{figure*}

The $^{238}$U($\gamma$,f) reaction cross section results are listed in Table \ref{table:ufission} and 
plotted in Fig. \ref{fig:uranium_contrast_cross_sections}, along with all available literature data sets which 
extend below an excitation energy of 6 MeV  
\cite{rabotnov1970photofission, ANDERL1973221, bowman863, ostapenko1978yields, zhuchko1978study, soldatov1992yield, manfredini1966results, KHAN1972333, MAFRA1972110, PhysRevLett.35.501, PhysRevC.21.1215, PhysRevC.87.044321} 
and the ENDF/B-VIII.0 evaluation \cite{BROWN20181}.

\begin{table}[h]
\centering
\begin{tabular}{ c  c  c }
\hline
\hline
E$_\gamma$ (MeV) \vline& $\sigma_{\text{E}_\gamma}$ (MeV) \vline& $\sigma(\gamma$,f) ($\mu$b) \\ \hline
4.3 & 0.065 & -0.3 $\pm$ 0.2\\
4.4 & 0.066 & -0.08 $\pm$ 0.1\\
4.5 & 0.068 & -0.06 $\pm$ 0.1\\
4.6 & 0.069 & 0.5 $\pm$ 0.1\\
4.7 & 0.071 & 1.2 $\pm$ 0.2\\
4.8 & 0.072 & 3.5 $\pm$ 0.2\\
4.9 & 0.074 & 8.9 $\pm$ 0.3\\
5.0 & 0.075 & 23.3 $\pm$ 0.2\\
5.1 & 0.077 & 37.0 $\pm$ 0.2\\
5.15 & 0.077 & 52.7 $\pm$ 0.2\\
5.2 & 0.078 & 88.0 $\pm$ 0.3\\
5.3 & 0.080 & 215.6 $\pm$ 0.6\\
5.6 & 0.084 & 2330 $\pm$ 5\\
5.7 & 0.086 & 2639 $\pm$ 6\\
5.8 & 0.087 & 2722 $\pm$ 6\\
5.9 & 0.089 & 3798 $\pm$ 7\\
6.0 & 0.090 & 7060 $\pm$ 10\\
 \hline
\end{tabular}
\caption{Tabulated $^{238}$U($\gamma$,f) reaction cross section data. The measurements were 
performed with Gaussian $\gamma$-ray beam spectra with mean E$_\gamma$ and spread 
$\sigma_{\text{E}_\gamma}$. The quoted errors are the statistical uncertainties, and there is an 
additional overall 3\% systematic error.}
\label{table:ufission}
\end{table}

The present results are in good agreement with the data of Ostapenko \textit{et al.} 
\cite{ostapenko1978yields}, Zhuchko \textit{et al.} \cite{zhuchko1978study} and 
Soldatov and Smirenkin \cite{soldatov1992yield} obtained using bremsstrahlung beams. Additionally, 
the present results are in remarkably good agreement 
with the ENDF/B-VIII.0 evaluation.

The present  $^{238}$U($\gamma$,f) reaction cross section data are consistent with the data of Csige 
\textit{et al.} \cite{PhysRevC.87.044321} above $E_\gamma\approx5.3$ MeV, however there is 
increasing disagreement between the data sets below this energy, with a factor of $\sim3$ discrepancy 
at 4.8 MeV. Because the data of Ref. \cite{PhysRevC.87.044321} was also obtained at the 
HI$\gamma$S facility, using nominally the same $\gamma$-ray beams as the present work, there is a unique 
opportunity to determine the source of the inconsistency between the prior measurement and the 
present results.
 We attribute this low energy divergence to a bremsstrahlung beam contamination 
present at the HI$\gamma$S facility, which was previously discussed in Ref. \cite{SCOTTCARMAN19961}. 
The effects of the bremsstrahlung beam were measured precisely for the first time in this work and were 
not taken into account in the results of Ref. \cite{PhysRevC.87.044321}. 
The $^{238}$U photofission cross section 
was measured in Ref. \cite{PhysRevC.87.044321} using an array of parallel plate avalanche 
counters, detecting both fragments from a fission event in coincidence. Since it is not possible to distinguish fission 
fragments from fission induced by the two different beam components, the background from the bremsstrahlung contamination of the 
HI$\gamma$S beam cannot be removed from the measured fission yields.

\begin{figure}[h]
\centering
\includegraphics[width=3.375in]{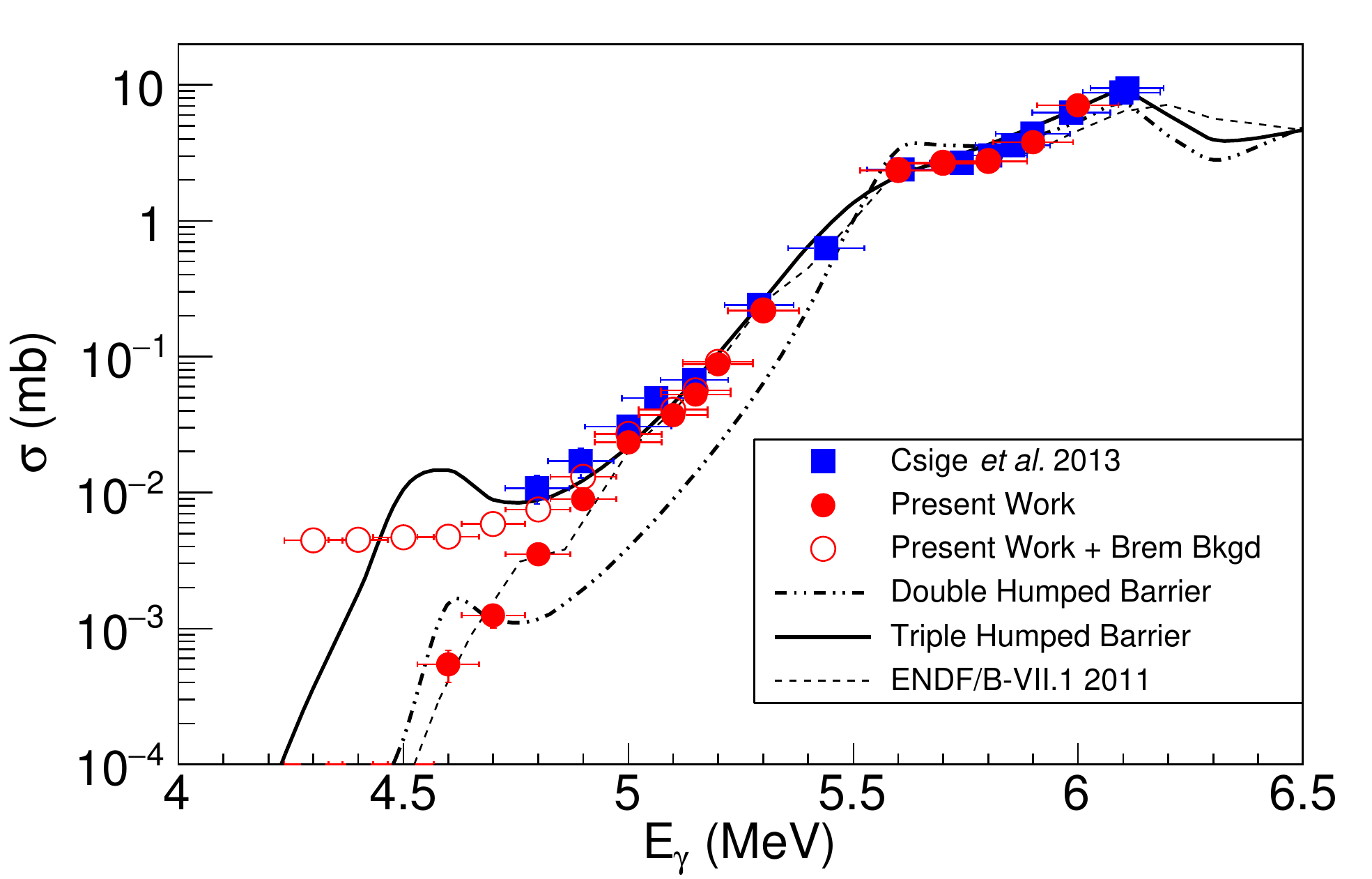}
\caption{(Color online) $^{238}$U photofission cross section results with and without the HI$\gamma$S 
$\gamma$-ray beam bremsstrahlung background subtracted, compared with the data of Csige 
\textit{et al.} \cite{PhysRevC.87.044321} and the ENDF/B-VIII.0 evaluation \cite{BROWN20181}. 
Photofission cross section calculations from Csige \textit{et al.} for double- and triple-humped fission 
barrier fits to their data are also shown. }\label{fig:csige_comparison}
\end{figure}

In Fig. \ref{fig:csige_comparison} the present results are shown with and without proper subtraction of 
the bremsstrahlung-induced background, with the non-background subtracted data in much better 
agreement with the data of Ref. \cite{PhysRevC.87.044321}. This agreement supports our assertion that the 
previously measured excess cross section at low energies was caused by the bremsstrahlung 
contamination of the HI$\gamma$S $\gamma$-ray beam.  Fig. \ref{fig:csige_comparison} also 
includes photofission cross sections calculated assuming double- and triple-humped $^{238}$U 
fission barriers by Csige \textit{et al}. The calculations were performed with the \textsc{empire 3.1} 
code \cite{HERMAN20072655}, tuning the input parameters to best reproduce the measured 
photofission cross section data. Neither calculation is in good agreement with the present results, in 
particular the predicted resonance at 4.6 MeV which is present in both calculations is not observed in 
the present data. The discrepancy between the current $^{238}$U photofission data and the 
calculated cross sections from Ref. \cite{PhysRevC.87.044321} assuming a double- and triple-
humped fission barrier casts significant doubt on the triple-humped shape of the $^{238}$U fission 
barrier proposed in Ref. \cite{PhysRevC.87.044321} and shown in Fig. \ref{fig:triple_humped_barrier}.

A more definitive statement about the structure of the $^{238}$U fission barrier will require new 
photofission cross section calculations tuned to fit not only the present data, but the broadest range of 
available data near the fission barrier. Given the good agreement between the present data obtained using
quasi-monoenergetic $\gamma$-ray beams and the data of Zhuchko \textit{et al.} \cite{zhuchko1978study}
obtained using bremsstrahlung beams, it 
is reasonable to conclude that the unfolding procedure applied to that data gives accurate results. 
Since the data of Ref. \cite{zhuchko1978study} extends down to E$_\gamma=3.42$ MeV, these data could 
supplement the present data set below E$_\gamma=4.6$ MeV, the lowest energy non-zero data point.

\section{Conclusions}

High-precision measurements of the photofission cross sections, prompt fission neutron 
polarization asymmetries, and the mean and spread of the prompt fission neutron multiplicity 
distributions of $^{232}$Th and $^{238}$U have been performed in the $\gamma$-ray energy region of E$_\gamma = 4.7-5.8$ MeV and 
E$_\gamma = 4.3-6.0$ MeV, respectively. This work, performed using the monochromatic, high-intensity, Compton-backscattered 
$\gamma$-ray beams of the HI$\gamma$S facility, represents the lowest energy measurements of 
this kind using quasi-monoenergetic $\gamma$-ray beams. 
Our results show that a previously observed shelf in the $^{238}$U photofission cross section, that had been identified as a resonance caused by a deep third minimum in the $^{238}$U fission barrier, 
is instead an accelerator background induced by a bremsstrahlung contamination of the $\gamma$-ray beam.
Future measurements of the photofission cross sections of both $^{238}$U and $^{232}$Th would help to place additional constraints on the structure of the fission barriers of the respective nuclei, especially with regards to the debate over the existence of deep third minima.
Next generation Compton-backscatter $\gamma$-ray sources such as the ELI-NP facility \cite{Thirolf_2012} and the proposed HI$\gamma$S2 upgrade \cite{higs2} will provide $\gamma$-ray beams with the improved flux and resolution needed to measure photofission cross sections and resonances well below the fission barrier.

\begin{acknowledgments}

The authors thank G. Rich for invaluable contributions to the development of this project.
Additionally, we express gratitude A. Banu and N. Parikh for their help in the experiment and data taking, and to J. Langenbrunner for providing material support.
We thank the HI$\gamma$S staff for delivering high-quality $\gamma$-ray beams for this work. We especially wish to thank S. Mikhailov for enlightening discussions of the $\gamma$-ray beam properties.
This work was performed under the auspices of US DOE by 
LLNL under contract DE-AC52-07NA27344, with support from the US DOE Office of Science, Office of Nuclear Physics through grants DE-FG02-97ER41033 and DE-FG02-97ER41041.

\end{acknowledgments}

\bibliography{photofission_paper}% Produces the bibliography via BibTeX.

\end{document}